\documentclass[aoas]{imsart}

\RequirePackage{amsthm,amsmath,amsfonts,amssymb}
\RequirePackage{natbib}
\RequirePackage[colorlinks,citecolor=blue,urlcolor=blue]{hyperref}
\RequirePackage{graphicx}
\usepackage{lmodern}
\usepackage[T1]{fontenc}
\usepackage{libertine}
 
\startlocaldefs
\theoremstyle{plain}

\theoremstyle{remark}

\endlocaldefs

\begin{document}

\begin{frontmatter}
\title{Bayesian Probit Multi-Study Non-negative Matrix Factorization for Mutational Signatures}
\runtitle{BaP Multi-NMF for Mutational Signatures}

\begin{aug}
\author[A]{\fnms{Blake}~\snm{Hansen}\ead[label=e1]{blake\_hansen@brown.edu}\orcid{0009-0004-3031-6409}},
\author[B]{\fnms{Isabella N.}~\snm{Grabski}\ead[label=e2]{igrabski@nygenome.org}\orcid{0000-0002-0616-5469}}
\author[C]{\fnms{Giovanni}~\snm{Parmigiani}\ead[label=e3]{gp@ds.dfci.harvard.edu}\orcid{0000-0002-8783-5961}}
\and
\author[A,D]{\fnms{Roberta}~\snm{De Vito}\ead[label=e4]{robeta\_devito@brown.edu}\orcid{0000-0003-0639-5341}}
\address[A]{Department of Biostatistics,
Brown University \printead[presep={,\ }]{e1,e4}}

\address[B]{New York Genome Center\printead[presep={,\ }]{e2}}

\address[C]{Department of Data Science, Dana Farber Cancer Institute \& Department of Biostatistics, Harvard T.H. Chan School of Public Health\printead[presep={,\ }]{e3}}

\address[D]{Data Science Institute, Brown University\printead[presep={,\ }]{e4}}
\end{aug}

\begin{abstract}

Mutational signatures are patterns of somatic mutations in tumor genomes that provide insights into underlying mutagenic processes and cancer origin. Developing reliable methods for their estimation is of growing importance in cancer biology.  Somatic mutation data are often collected for different cancer types, highlighting the need for multi-study approaches that enable joint analysis in a principled and integrative manner. Despite significant advancements, statistical models tailored for analyzing the genomes of multiple cancer types remain underexplored.
In this work, we introduce a Bayesian Multi-Study Non-negative Matrix Factorization (NMF) approach that uses mixture modeling to incorporate sparsity in the exposure weights of each subject to mutational signatures, allowing for individual tumor profiles to be represented by a subset rather than all signatures, and making this subset depend on covariates. This allows for a) more precise ability to identify meaningful contributions of mutational signatures at the individual level; b) estimation of the prevalence of activity of signatures within a cancer type, defined by the proportion of tumor profiles where a certain signature is present; and c) de-novo identification of interpretable patient subtypes based on the mutational signatures present within their mutational profile. We apply our approach to the mutational profiles of tumors from seven different cancer types, demonstrating its ability to accurately estimate mutational signatures while uncovering both individual and tissue-specific differences. An R package implementing our method is available at \href{https://github.com/blhansen/BAPmultiNMF}{github.com/blhansen/BAPmultiNMF}.
\end{abstract}

\begin{keyword}
\kwd{Mutational Signatures Analysis}
\kwd{Bayesian Multi-Study NMF}
\kwd{Latent Variable Modeling}
\end{keyword}

\end{frontmatter}

\section{Introduction}

Somatic mutations play a central role in cancer progression, accumulate in cell genomes, and drive tumorigenesis~\citep{Vogelstein2004,loeb_significance_2000, stratton_cancer_2009, stratton_exploring_2011}. Somatic mutation patterns, known as mutational signatures, are important as they can reflect the mutagenic processes responsible for these genomic alterations~\citep{nik-zainal_mutational_2012, alexandrov_repertoire_2020}. Identifying and characterizing these signatures has emerged as an effective way to uncover the molecular mechanisms underlying cancer development and progression~\citep{alexandrov_signatures_2013, helleday_mechanisms_2014}. Mutational signatures have been associated with the exposure of each subject to both environmental and endogenous factors, including to mutagenic agents such as tobacco smoke, ultraviolet (UV) light, and deficiencies in DNA repair mechanisms~\citep{nik-zainal_mutational_2012, alexandrov_signatures_2013, hayward_whole-genome_2017}. More recently, mutational signatures have been used to guide precision medicine by linking specific signatures to the mechanisms behind therapeutic interventions~\citep{Gulhan2019,samur_genome-wide_2020,koh_mutational_2021,Jin2024}. 

Mutational signatures present an opportunity to understand heterogeneity within cancer types by detecting which mutational signatures are present at the subject level, providing greater insight into individual cancer genomes than can be provided by broadly comparing cancer types. To accomplish this goal, we analyzed mutational profiles from a collection of seven cancer types that contain several cancers known to be caused by tobacco smoking. These include cancers with strong associations with smoking, such as lung squamous cell carcinoma and lung adenocarcinoma, and some cancers with a link to smoking only in a subset of patients, such as head squamous cell carcinoma. Our data is obtained from the Pan-Cancer Analysis of Whole Genomes (PCAWG), a comprehensive resource containing whole genome sequencing data from 2,780 cancer genomes across 37 distinct tumor types~\citep{aaltonen_pan-cancer_2020}. This dataset offers an excellent opportunity to investigate mutational processes, due to its extensive scope and the inclusion of clinical and demographic information. Understanding the set of mutational signatures present within a related group of cancers not only enhances our understanding of the molecular mechanisms underlying cancer development, but also has the potential to guide personalized treatment strategies. Ultimately, this research contributes to the broader objective of understanding the causes of cancer by identifying and understanding relevant mutational processes.

To systematically uncover the mutational processes driving cancer, mathematical and statistical methods have been extensively developed to identify mutational signatures from profiles of somatic mutations. Among these approaches, Non-Negative Matrix Factorization (NMF) has emerged as a powerful and widely adopted technique for decomposing mutational profiles into a matrix of mutational signatures and a matrix of exposures, which quantifies the exposure weights of each subject to mutational signatures within each tumor~\citep{nik-zainal_mutational_2012, alexandrov_signatures_2013}. Several computational tools, such as SigProfilerExtractor~\citep{islam_uncovering_2022} and MuSiCal~\citep{Jin2024}, have implemented NMF-based frameworks to enable robust mutational signature analysis. Similar tools have also been developed using a Bayesian perspective, such as signeR~\citep{rosales_signer_2017} and compressive NMF~\citep{zito_compressive_2024}. A broad family of alternatives, with some overlap, are based on Hierarchical Dirichlet Processes~\citep{roberts_patterns_2018}. 

Despite these advancements, most existing methods are what we refer to as single-study approaches, meaning they are intended to analyze one dataset at a time. For example, to extract mutational signatures from samples of multiple cancer types, these methods would either analyze each cancer type separately, requiring extensive and heuristic pre- and post-processing steps to synthesize results, or stack cancer types into a single dataset under the assumption of a homogeneous data-generating process. Both approaches have limitations: analyzing each cancer type separately is inefficient and cannot borrow strength across cancer types to estimate shared mutational signatures, a critical downside especially when the number of samples per cancer type is small. At the same time, stacking does not account for the heterogeneity inherent in cancers of different types, potentially compromising the estimation of shared or cancer-type-specific components~\citep{de_vito_multistudy_2019}. These ideas extend to other cases where multiple datasets might be analyzed, such as when considering data from multiple studies or populations. Given the complexity and lack of large sample datasets available in cancer genomics, there is a clear need for what we refer to as multi-study methods that can effectively capture both shared and study-specific features.

Multi-study methods \citep{roy_perturbed_2021, grabski2023bayesian, chandra2024inferring} provide a principled framework for addressing these challenges by explicitly modeling that data arise from distinct studies, groups, or conditions. For simplicity, we refer to a member of any grouping structure as a study. These methods enable the pooling of data to estimate shared signals while capturing study-specific variations. However, most of these methods are designed to analyze continuous data that can be modeled with a normal distribution, whereas mutational counts require alternative distributional assumptions. In the context of cancer genomics, a multi-study framework can be used to treat each cancer type as a separate study, allowing for efficient estimation of shared signal while retaining cancer-type-specific nuances. Multi-study frameworks could also be applied to analyze multiple samples of the same cancer type under different conditions, such as treating subjects with different ages, smoking histories, geographic locations, or therapies as different studies. The recently developed Bayesian Multi-Study Non-Negative Matrix Factorization (Ba Multi-NMF)~\citep{grabski_bayesian_2023} introduced a Bayesian framework to jointly analyze mutational count data across multiple types of cancer. Although Ba Multi-NMF marked a significant advancement, it has key limitations: (1) signatures are represented as present or not in each study using study-level binary indicators, limiting the ability to characterize signature prevalence within a study population; (2) covariates cannot be directly used to estimate the presence of mutational signatures at the study-level; and (3) reliance on computationally intensive Markov Chain Monte Carlo (MCMC) methods limits scalability to larger datasets.

In this paper, we address the key limitations of the Bayesian Multi-Study NMF model proposed by ~\cite{grabski_bayesian_2023} and introduce a novel model that incorporates significant advancements. A central innovation of our method lies in its ability to estimate sparse exposures of mutational signatures at the individual sample level.   Unlike the previous approach, which relied on binary presence/absence indicators at the study level, our model includes binary presence/absence indicators at the individual level, enabling the assessment of the presence of an exposure for each sample. This allows for a) increased sparsity on the exposure matrix; b) more precise ability to identify meaningful contributions of mutational signatures at the individual level; c) estimation of the prevalence of activity of signatures within a study, defined by the proportion of cancer genomes where a certain signature is present; and d) de-novo identification of interpretable patient subtypes based on the presence/absence indicators for one or more signatures. Additionally, our method can incorporate covariates' information through a mixture prior facilitating the discovery of dependencies between subject specific risk factors and the presence of mutational signatures, enabling more effective and more interpretable clustering of samples compared to other NMF-based methods, and paving the way for future understanding into the molecular mechanisms behind signatures. Computationally, an important innovation of this work is the use of Variational Bayesian inference, which significantly reduces the computational burden and enables efficient analysis of large-scale datasets.

The paper proceeds as follows. Section 2 describes the seven types of cancer that we analyze from the PCAWG dataset. Section 3 defines the Bayesian Probit Multi-Study NMF model and prior structure.  Section 4 presents simulation studies comparing our model to the existing Bayesian Multi-Study NMF framework.  Section 5 applies Bayesian Probit Multi-Study NMF to the mutational data we introduce in Section 2 and discusses the key findings. Finally, Section 6 provides a discussion and directions for future work.

\section{Pan Cancer Analysis of Whole Genomes (PCAWG) Case study}

In this work, we focus on analyzing mutational signatures derived from single-base substitution (SBS) mutational profiles~\citep{alexandrov_signatures_2013, alexandrov_repertoire_2020}. SBS mutations occur when one DNA nucleotide is replaced by another, and they are classified based on the modified pyrimidine of the mutated nucleotide base pair, resulting in six possible base substitutions: C>A, C>G, C>T, T>A, T>C, and T>G. It is typical to define the mutation signature by also considering the two adjacent nucleotides, which leads to a total of 96 unique mutation motifs, although other alphabets can also be considered. The data for each subject is the number of mutations of each motifs across their sequenced cancer genome. This provides a summary of the mutational landscape of the tumor relative to the patient's germline genome.

Our data comes from mutational profiles in the PCAWG dataset~\citep{aaltonen_pan-cancer_2020}, accessible through the XENA platform~\citep{goldman_visualizing_2020}. The PCAWG dataset encompasses mutational profiles from 2,780 cancer genomes across 37 distinct tumor types, representing one of the most comprehensive resources for studying mutational processes in cancer. We focus on seven types of cancer and associated covariates summarized in Table~\ref{tab:the_studies}. This table provides an overview of the demographic and clinical characteristics. Cancer types vary in terms of the proportion of male subjects and the average age at diagnosis. This information highlights the variability in sample size, sex distribution, and age demographics across cancer types, reflecting the diversity and complexity of the dataset.

We excluded certain covariates for specific cancer types in our analysis due to data limitations. For Breast Adenocarcinoma, the sex covariate was omitted due to extreme class imbalance, while both sex and age covariates were excluded for Esophageal Adenocarcinoma due to insufficient complete cases (18 complete cases out of of 98 profiles). Consequently, we analyzed the $N_s=98$ mutational profiles for this cancer type without incorporating any covariates. The remaining cancer types were analyzed with complete mutational profiles and available covariates.

\begin{table}[ht]
    \centering
    \begin{tabular}{l c c c}
    \hline
       Cancer Type  &  $N_s$ & Proportion Male Sex & Age at Diagnosis \\
       \hline 
       Breast Adenocarcinoma  & 177(198) & \textcolor{red}{0.05\%}& 56(13)\\
       Colorectal Adenocarcinoma & 58(60) & 50\% & 65(8)\\
       Esophageal Adenocarcinoma & \textcolor{red}{18(98)} & \textcolor{red}{83\%} & \textcolor{red}{70(8)}\\
       Head Squamous Cell Carcinoma & 56(57) & 82\% & 53(14) \\
       Lung Adenocarcinoma & 35(38) & 47\% &  65(10) \\
       Lung Squamous Cell Carcinoma & 48(48) & 79\% & 66(9)\\
       Stomach Adenocarcinoma & 51(75) & 71\% & 65(12) \\
     \hline
    \end{tabular}
    \caption{Summary of PCAWG case study data. For each cancer type, the number of subjects (complete cases(total)), proportion of male subjects, and mean (standard deviation) age at diagnosis are provided. Covariate exclusions were made for Breast Adenocarcinoma and Esophageal Adenocarcinoma due to data limitations.}
    \label{tab:the_studies}
\end{table}

Age provides a useful example of how the use of covariates can increase both efficiency and interpretability. Mutations accumulate in somatic cells before the onset of cancer \cite{Tomasetti2013}.  
Past research has identified specific signatures for mutations more commonly associated with age at diagnosis \cite{alexandrov_repertoire_2020}. 
Thus, the cancer genomes of patients with older ages at diagnosis are more likely to present with these signatures. Explicitly including the age covariate will increase the model's ability to assess the presence of these signatures, and will enable the model to estimate these signatures mostly using older individuals, making it more specific.

\section{Bayesian Probit Multi-Study Non-negative Matrix Factorization}

\subsection{Model Specification}\label{sec:mod_spec}
Consider mutational profiles data from $S$ different studies. For each study $s$, we observe $\mathbf{M}_s$, a counts matrix of dimension $K \times N_s$, where $K$ is the dimension of the mutational profiles and $N_s$ is the number of subjects. In the context of SBS mutational profiles, $K=96$ unique mutation motifs are defined, although the framework applies to any counts data with $K\in\mathcal{Z}^+$ categories.

Assume there exist matrices $\mathbf{P}$, $\mathbf{E}_s$ and $\mathbf{W}_s$ such that:
\begin{align}
    \mathbf{M}_s &\sim \text{Poisson}\left(\mathbf{P}\mathbf{E}_s \mathbf{W_s}\right),\label{multi-study-alloc-nmf} \;\;\;s=1,\ldots,S
\end{align}
where $\mathbf{P}$ is the $K \times R$ \textit{signatures} matrix, $\mathbf{E}_s$ is the $R \times N_s$ \textit{exposures} matrix, and $\mathbf{W_s}=\text{diag}\left(w_{s1},\ldots,w_{sN_s}\right)$ are subject-specific \textit{weights}.
For a specific entry, the generative model can be written as:
 \begin{eqnarray}
   m_{sij} &\sim \text{Poisson}\left(w_{sj}\sum_r p_{ir}e_{sjr}\right),  
 \end{eqnarray}
where $s=1,\ldots,S$, motif $i=1,\ldots,K$, signature $r=1,\ldots, R$, and subject $j=1,\ldots,N_s$, 
 
To facilitate computation, we augment this model with latent variables $\mathbf{Z}_s\in\mathcal{R}_+^{K \times R \times N_s}$:
\begin{align}
    z_{sijr} &\sim \text{Poisson}\left(p_{ir}e_{sjr}w_{sj}\right) \label{z_poisson_mean}.
\end{align}
Here, $z_{sijr} $ represents the latent count of motif $i$ attributed to signature $r$ for subject $j$ in study $s$.

\subsection{Prior Specification}\label{model_priors_basic}
We specify the following priors for the model parameters:
\begin{align}
    \mathbf{P}_{r} &\sim \text{Dirichlet}\left(\boldsymbol{\alpha}^{p}\right), \label{prior_p_dir}\\
    \mathbf{E}_{sj} &\sim \text{Dirichlet}\left(\boldsymbol{\alpha}^{e}_{sj}\right),\label{prior_e_dir} \\
     w_{sj} &\sim \text{Gamma}\left(\alpha^{w}_s, \beta^{w}_s\right),\label{prior_weight}
\end{align}
where $\mathbf{P}_{r}$ is column $r$ of $\mathbf{P}$, $\mathbf{E}_{sj}$ is column $j$ of $\mathbf{E}_{s}$, and $\alpha^{w}_s,\beta^{w}_s\in\mathcal{R}_+$
are shape and rate parameters for the Gamma distribution. We place hyperpriors on the Gamma parameters:
\begin{align}
    \alpha^{w}_s &\sim \text{Exp}\left(\lambda_s^w\right),\label{prior_w_shape} \\
    \beta^{w}_s &\sim \text{Gamma}\left(a_s^w,b_s^w\right),\label{prior_w_rate}
\end{align}
where $\lambda_s^w, a_s^w, b_s^w \in \mathcal{R}_+$.

The concentration parameters $\boldsymbol{\alpha}^{p} \in \mathcal{R}^{K}_+$, $\boldsymbol{\alpha}^{e}_s\in \mathcal{R}^{R}_+$, are treated as hyperparameters, with further specifications detailed in Section~\ref{exposure_prior_section}.

Under priors~\eqref{prior_p_dir}-\eqref{prior_weight}, the columns of $\mathbf{P}$ and $\mathbf{E}_s$ are normalized, ensuring the columns of $\mathbf{P}\mathbf{E}_s$ sum to 1~\citep{yildirim_bayesian_2021}. An intuitive interpretation of Bayesian Multi-Study NMF is that the observed mutational profile $\mathbf{M}_{sj}$ is explained by allocating $w_{sj}$ total mutations across the $K$ mutation motifs according to latent mutational signatures $\mathbf{P}$ and subject-specific weights $\mathbf{E}_s$. This property also simplifies likelihood computations and mitigates the scaling unidentifiability typical of NMF.

\subsection{Latent Probit Regression Model}\label{exposure_prior_section}

To incorporate covariate information into the mutational signature selection process, we extend the prior of $\mathbf{E}_s$ with a probit hyper-prior structure. Specifically, we use a mixture prior on the concentration parameters  $\alpha_{sjr}$:
\begin{align}
    \mathbf{E}_{sj} &\sim \text{Dirichlet}\left(\left\{(a_{sjr})\alpha^{e1}_s + (1-a_{sjr})\alpha^{e0}_s\right\}_{r=1}^R\right),\label{prior_dirichlet_mixture}
\end{align}
where $a_{sjr}$ is a binary indicator:
    \begin{align}
    a_{sjr} &=\begin{cases} 
      1 & a^*_{sjr} > 0 \\
      0 & a^*_{sjr} \leq 0
   \end{cases}, 
 \end{align}  
   and 
\begin{align}   
   a^*_{sjr} &\sim \mathcal{N}\left(\boldsymbol{\beta}_{sr}^\intercal\boldsymbol{x}_{sj},1\right), \;\;\;\;  \boldsymbol{\beta}_{sr} \sim \mathcal{N}\left(\boldsymbol{\beta}_0,\tau_{sr}^{-1}\boldsymbol{I}_Q\right), \;\;\;\;
   \tau_{sr} \sim \Gamma\left(\gamma_1, \gamma_2\right).\label{beta_var_prior}
\end{align}
Here, $\boldsymbol{x}_{sj}$ is the $Q$-dimensional vector of covariates for subject $j$ in study $s$.
Thus, the mixture is defined by latent indicator $a_{sjr}$ which depends on the latent variable $a^*_{sjr}$, whose distribution is influenced by covariates through study- and signature-specific regression coefficients $\boldsymbol{\beta}_{sr}$. This construction allows covariates to inform the allocation of mutational signatures at the subject level, while still allowing for inference about the presence of signatures at the study level.

\subsection{Parameter estimation: CAVI algorithm}
We employ a Coordinate Ascent Variational Inference (CAVI) algorithm to estimate the model parameters. CAVI optimizes a predefined family of approximate posteriors to minimize the Kullback-Leibler divergence from the true posterior distribution~\citep{bishop_pattern_2006}. Compared to MCMC-based methods, CAVI offers computational efficiency while maintaining accuracy in parameter estimation~\citep{hansen_paper}. To address sensitivity to initialization, we provide multiple initialization strategies, ranging from  random initialization to leveraging frequentist methods. Full details of the CAVI algorithm and initialization strategy are provided in Supplementary Section 2.

\subsection{Recovery-Discovery Prior}
To incorporate prior knowledge, such as previously identified mutational signatures, we adopt a recovery-discovery framework.
Specifically, \cite{grabski_bayesian_2023} and \cite{zito_compressive_2024} both include versions of their Bayesian NMF models that incorporate mutational signatures from the COSMIC database~\citep{tate_cosmic_2019} through informative priors. 

Let $\boldsymbol{\Gamma}^{Recov}$ represent previously known signatures.  We modify our BaP Multi-NMF model as:
\begin{equation}
    \mathbf{M}_s \sim \text{Poisson}\left(\mathbf{P}^{Recov}\mathbf{E}^{Recov}_s \mathbf{W_s}+\mathbf{P}^{Discov}\mathbf{E}^{Discov}_s \mathbf{W_s}\right),\label{multi-study-alloc-nmf-rd}
\end{equation}
where $\mathbf{P}^{Recov}$ and $\mathbf{E}^{Recov}_s$ represents known signatures and their exposures, while   $\mathbf{P}^{Discov}$ and $\mathbf{E}^{Discov}_s$ represent \textit{de novo} signatures. The priors for $\mathbf{E}^{Recov}_s$, $\mathbf{P}^{Discov}$, $\mathbf{E}^{Discov}_s$, and $\mathbf{W_s}$ are defined in Sections \ref{model_priors_basic}-\ref{exposure_prior_section}, while prior for $\mathbf{P}^{Recov}$ is defined as:
\begin{equation}
    \mathbf{P}^{Recov}_r \sim \text{Dirichlet}\left(c_r \boldsymbol{\gamma}^{Recov}_r\right),
\end{equation}
where $c_r\gg0$, $r=1,\ldots,R_{recov}$, $\boldsymbol{\gamma}^{Recov}_r$ is signature $r$ of $\boldsymbol{\Gamma}^{Recov}$. This prior, a generalized multi-study version of \cite{zito_compressive_2024}, uses previously discovered signatures multiplied by a large constant as the Dirichlet concentration parameter. 

This approach directly incorporates prior knowledge while allowing the discovery of novel signatures and their dependencies.
Specifically, since BaP Multi-NMF directly uses the recovered signatures in the Dirichlet concentration parameter, the incorporation of new canon mutational signatures or in any general NMF setting is more easily implemented in practice compared to the procedure proposed by \cite{grabski_bayesian_2023}, which involves generating synthetic data based on the signatures of interest and using the resulting posterior distribution over $\mathbf{P}^{Recov}$ as the new prior in subsequent analyses. 

An additional advantage of our approach is that we can combine the recovery-discovery approach in a multi-study setting with the latent probit regression model described in Section~\ref{exposure_prior_section}, allowing us to detect relevant mutational signatures even within subgroups in each study. This is an improvement over \cite{grabski_bayesian_2023}, which can only detect study-level presence or absence of mutational signatures.


\section{Simulation Studies}\label{sec:sims}

\begin{figure}[ht]
    \centering
    \includegraphics[width=\linewidth]{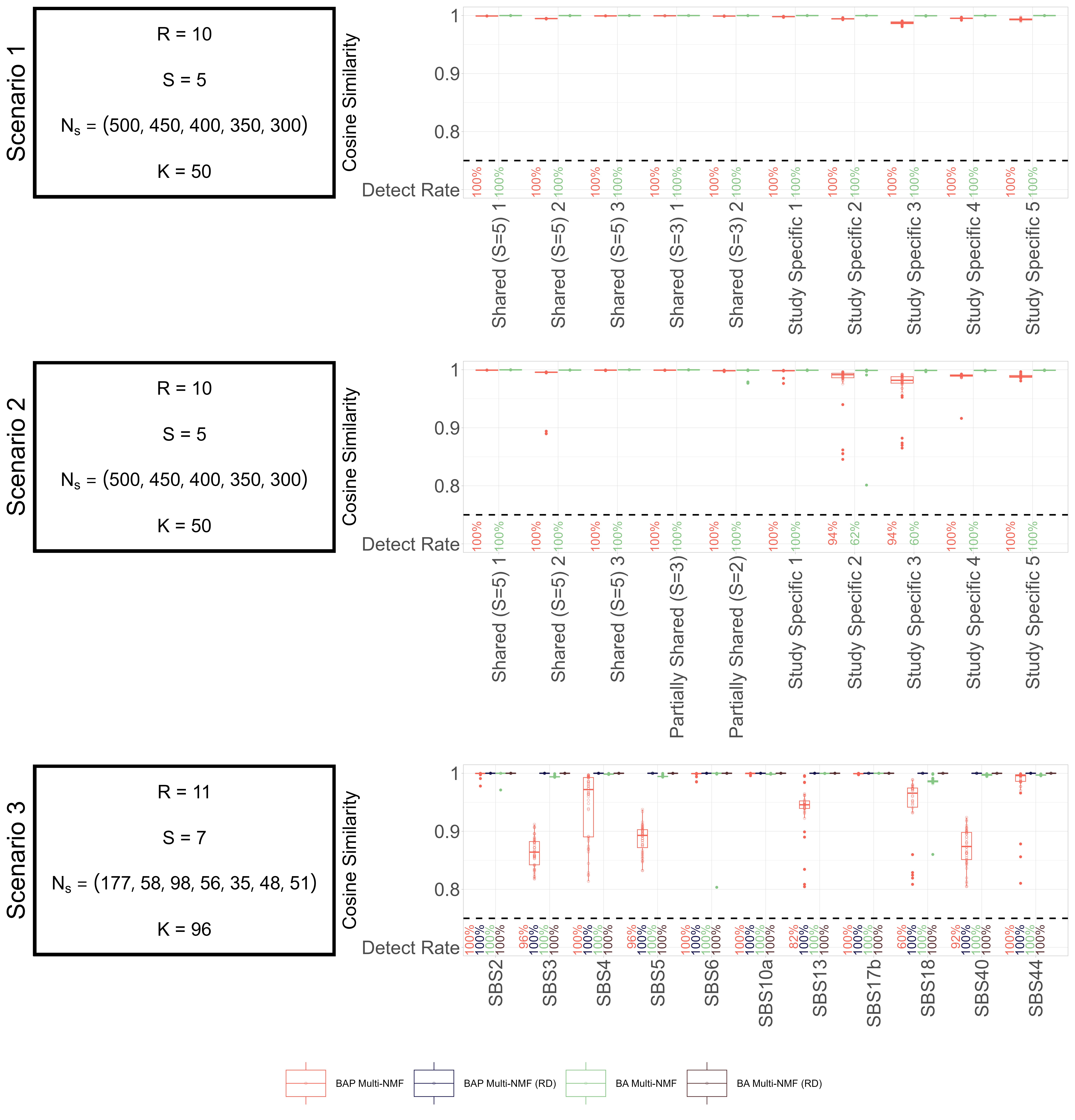}
    \caption{Summary of simulation scenarios. The first column describes simulation settings. The second column shows the distribution of cosine similarities between the estimated and ground truth signatures for BaP Multi-NMF and Ba Multi-NMF. A signature was considered captured if its highest cosine similarity to any estimated signature was above 0.8, otherwise the signature was considered missed. We also report the detection rate for each signature, calculated as the proportion of simulations that captured a given signature.}
    \label{fig:cos_sim_plot}
\end{figure}

In this section, we present a series of simulation studies to evaluate the performance of the BaP Multi-NMF model in estimating mutational signatures and assigning them at both the study and individual levels. We considered three distinct scenarios to assess the model’s performance (Figure~\ref{fig:cos_sim_plot}, 1st column). In Scenario 1, the presence of signatures are binary, and without influence from the covariates. Scenario 2 extends Scenario 1 by incorporating several covariates that affect signature presence, which allowed us to simulate scenarios where some mutational signatures are present within samples with a small probability. This scenario aimed to mimic more realistic conditions, where some mutational signatures are only present across a small subset of individuals or studies in a covariate-specific manner. Scenario 3 was based on real data derived from the the Pan-Cancer Analysis of Whole Genomes (PCAWG) dataset, offering a robust assessment of the model’s ability to detect known mutational signatures.

We compared BaP Multi-NMF to Ba Multi-NMF~\citep{grabski_bayesian_2023}, available at \href{https://github.com/igrabski/MultiStudyNMF}{igrabski/MultiStudyNMF}, which is currently the only multi-study NMF tool developed for mutational signature analysis. We utilized the \texttt{discovery\_fit} and \texttt{recovery\_discovery\_fit} functions of Ba Multi-NMF with default parameters.

\subsection{Simulation Parameters}
\begin{figure}
    \centering
    \includegraphics[width=\linewidth]{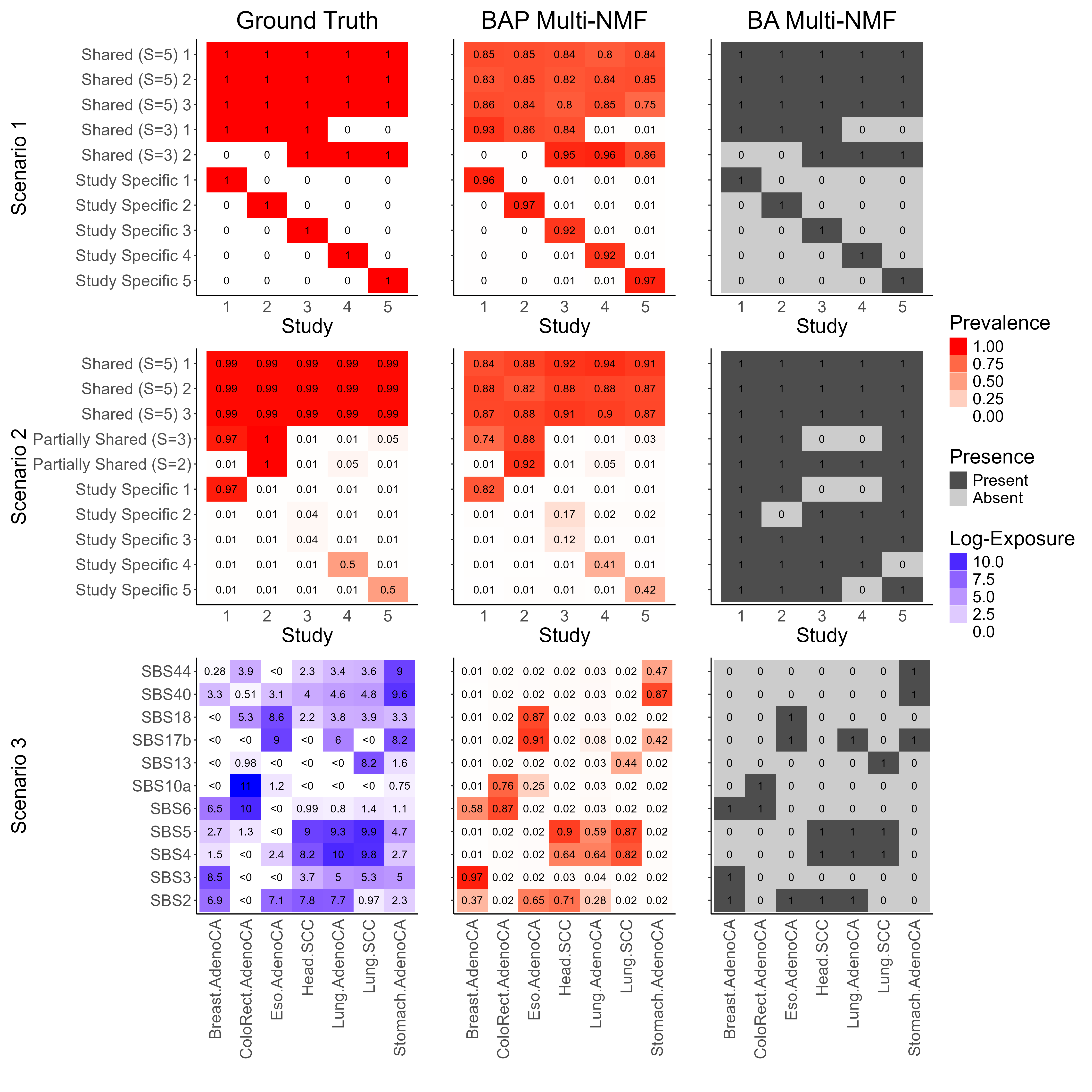}
    \caption{Signature assignment results across the three simulation scenarios. For the ground truth and BaP Multi-NMF results, the average of the ground truth and estimated probabilities of the binary indicator $a_{sjr}$ was computed as $1/N_s\sum_{j=1}^{N_s}\Phi\left(\boldsymbol{\beta}_{sr}^\intercal\boldsymbol{x}_{sj}\right)$. In Scenario 3, where ground truth inclusion probabilities are not available, the average log unnormalized exposure were reported. For Ba Multi-NMF, which assigns a binary indicator for each study ($a_{sj}$), the mode across the 50 simulation replicates was used.}
    \label{fig:inc-prob}
\end{figure}

In Scenario 1, the signature matrices $\mathbf{P}$ were generated from a Dirichlet distribution with a constant concentration parameter of $0.1$. Exposures were sampled from a gamma distribution with shape $2$ and rate $10$, and subject-specific weights were set to $w_{sj}=1000$. Scenario 2 carries the same parameters as Scenario 1, except subjects were modeled as having non-zero exposures with probabilities determined by a probit model using covariates, i.e., $\Phi\left(\sum_{i=0}^4 \beta_i x_i \right)$, where $x_0=1$ acts as an intercept, $x_1 \sim \text{Bernoulli}(0.2)$ is a binary covariate, and both $x_2,x_3\sim \mathcal{N}(0,1)$ are continuous covariates. This approach allowed simulation of scenarios where some signatures were associated with specific covariate profiles.  The average inclusion probabilities for this scenario are summarized in  Figure~\ref{fig:inc-prob}. 

To generate ground truth parameters in Scenario 3, single-study frequentist NMF models were fitted to each of the seven cancer types listed in Table~\ref{tab:the_studies} and the resulting mutational signatures were matched to those in the COSMIC v3.2 database~\citep{tate_cosmic_2019}, using a cosine similarity threshold of $\geq 0.7$. The matched COSMIC signatures ($\mathbf{P}_{COSMIC}$) were used as the ground truth, and unnormalized exposures ($\mathbf{E}_s$) were estimated by minimizing the Frobenius norm $|| \mathbf{M}_s - \mathbf{P}_{COSMIC}\mathbf{E}_s ||_{F}$. We generate the ground truth parameters for Scenario 3 using this process because it reflected the complexity of real-world data and served as a robust assessment of the model’s ability to detect known mutational signatures.

Fifty simulation replicates were generated for each scenario by sampling from Model~\eqref{multi-study-alloc-nmf}. Performance was evaluated by comparing the estimated signatures to the ground truth using cosine similarity.

\subsection{Estimation Accuracy}

Figure~\ref{fig:cos_sim_plot} illustrates the estimation accuracy of mutational signature recovery for BaP Multi-NMF and Ba Multi-NMF across the three scenarios. When the highest cosine similarity between a ground truth signature and estimated signature is below 0.8, we consider that signature to have been missed by the model; we report the proportion of simulations that capture signatures as the detection rate. In Scenario 1, where the total sample size was relatively high ($\sum N_s = 1700$ vs. $K = 50$), both BaP Multi-NMF and Ba Multi-NMF achieved high accuracy, with average cosine similarity exceeding 0.99 and a 100\% detection rate for all signatures. Scenario 2, where signatures were expressed at lower levels, revealed key differences between the methods. Ba Multi-NMF showed reduced detection rates for certain study-specific signatures, specifically signatures 2 and 3, with detection rates of 62\% and 60\%, respectively. In contrast, BaP Multi-NMF maintained high detection rates ($\geq 94\%$) due to its ability to model signature assignment at the individual level. These results highlight the differences between Ba Multi-NMF and BaP Multi-NMF: Ba Multi-NMF treats signature assignment as binary and at the study level, when BaP Multi-NMF models signature assignment at the subject-level using mixtures, enabling the detection of signatures which are present in a small subset of samples within a study without having to commit to a binary indicator at the study level. Scenario 3, with a smaller total sample size ($\sum N_s = 523$) and higher motifs dimension ($K=96$), posed greater challenges. In this scenario, BaP Multi-NMF’s discovery-only approach maintains a high accuracy, with the average cosine similarity $\geq 0.8 $, while Ba Multi-NMF consistently achieved an average cosine similarity of 0.99. However, the recovery-discovery versions of both models performed equally, achieving a similarly high average cosine similarity, i.e. $\geq 0.99$. These findings demonstrate that incorporating prior knowledge through the recovery-discovery priors mitigates potential limitations of BaP Multi-NMF in small-sample size settings.

\subsection{Signature Assignment Accuracy}

The ability of BaP Multi-NMF to correctly assign signatures at the study level compared to Ba Multi-NMF is summarized in Figure~\ref{fig:inc-prob}. In Scenario~1, where signatures were universally present or absent, both BaP Multi-NMF and Ba Multi-NMF were able to accurately assign the signatures across each study, replicating the ground truth signature-sharing pattern. In Scenario~2, where some signatures are expressed at lower proportions, BaP Multi-NMF accurately estimated the proportion of individuals exposed to each signature, even for low-prevalence signatures (e.g., Signature 4 with a ground truth prevalence of 5\% in Study 5). In contrast, Ba Multi-NMF, which assigns binary indicators at the study level, can only assign an indicator of 1 to signatures, even if they are only present within 1\% of samples. Additionally, for rare signatures, Ba Multi-NMF has a much lower detection rate - 60-62\% compared to BaP Multi-NMF which detected both signatures in 94\% of simulations. In Scenario 3, where real-data-inspired parameters were used, the ground-truth parameters are modeled after real data and therefore there are no ground-truth inclusion probabilities to assess. Instead, we report the average log (unnormalized exposures), which relate to the number of mutations attributable to a mutational signature. An important difference from Scenarios 1 and 2 is that in this scenario, the mutational burden imposed by each signature can vary in magnitude. Compared to the true exposures, both BaP Multi-NMF and Ba Multi-NMF accurately detected dominant signatures, but missed low-prevalence signatures responsible for significantly fewer mutations. For example, both methods only detect SBS44 in Stomach Adenocarcinoma, even though it is responsible for about 20-50 mutations on average in most of the other studies. 
Overall, these results suggest that while both BaP Multi-NMF and Ba Multi-NMF can accurately detect influential mutational signatures, BaP Multi-NMF provides higher resolution in signature assignment by working at the subject level, allowing for more detailed comparisons across cancer types. For instance, BaP Multi-NMF assigns an average assignment probability of 0.64 and 0.82 to SBS4 in Lung adenocarcinoma and Lung squamous cell carcinoma, respectively, indicating the signature’s importance in these studies in a more granular manner than would be possible with binary indicators. 

\begin{figure}
    \centering
    \includegraphics[width=\linewidth]{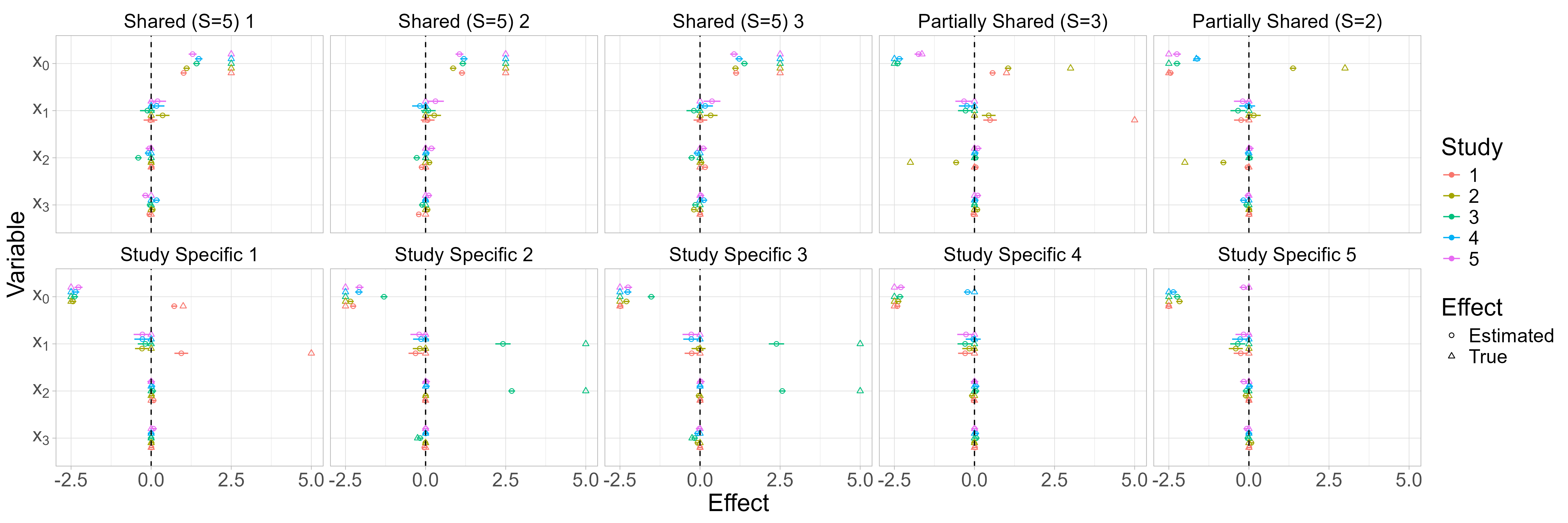}
    \caption{BaP Multi-NMF signature assignment results for scenario 2. For each effect, we report a 95\% credible interval based on the average mean and covariance of $\boldsymbol{\beta}_{sr}$ across all 50 simulations. We note that $x_0 = 1$ acts as an intercept term and can be interpreted as a study-level prevalence conditional that $x_1=0$, $x_2=\bar{x}_2$ and $x_3=\bar{x}_3$.}
    \label{fig:scen2-cov-plot}
\end{figure}

In addition to study-level prevalence, BaP Multi-NMF can assess the association between signature assignment at the subject level and measured covariates. Figure~\ref{fig:scen2-cov-plot} displays the estimated covariate effects compared to the ground-truth values. Overall, we find that BaP Multi-NMF effectively recovers the correct relationship between the covariates and mutational signatures, though it may under-estimate the effect magnitude. This is because the signature assignment process assigns $\hat{a}_{sjr} = 0$ when $e_{sjr}$ is near 0, even when the true $a_{sjr} = 1$ with a small non-zero $e_{sjr}\ll 0.01$. Despite this, BaP Multi-NMF consistently identifies the correct direction of covariate effects.  BaP Multi-NMF successfully identified Study Specific Signature 3 as present in samples from Study 3 with $x_1=1$ and above average $x_2$ values, with a slight negative correlation with $x_3$. These findings demonstrate BaP Multi-NMF’s ability to accurately associate covariates with the presence of mutational signatures.

\subsection{Computational Efficiency}

Finally, computational efficiency was evaluated, with results shown in Supplementary Figure S1. The computational time and memory for both methods were monitored using the \texttt{peakRAM} R package~\citep{peakRAM}. Across all scenarios, BaP Multi-NMF demonstrated significant computational advantages over Ba Multi-NMF. This efficiency arises because  BaP Multi-NMF employs Variational Bayes (VB) for estimation,  whereas Ba Multi-NMF relies on a hybrid Gibbs Sampler with a long tempering phase (full details at ~\cite{grabski_bayesian_2023}). In Scenarios 1 and 2, BaP Multi-NMF completed simulations in an average of 6 minutes and 250 MB of RAM. In Scenario 3, BaP Multi-NMF's discovery-only version required 22 minutes and 69 MB of RAM, while Ba Multi-NMF needed 5.5 hours and 3 GB. When employing the recovery-discovery versions, BaP Multi-NMF completed the analyses in 1 hour using 137 MB of RAM,  whereas Ba Multi-NMF’s  required 10 hours and 7.6 GB. These findings underscore  the significant computational efficiency of the variational approach implemented in BaP Multi-NMF, making it particularly well-suited for analyzing large-scale mutational signature datasets as they become increasingly available.

\section{Mutational signatures analysis}

In this section, we analyze the collection of cancer types from the PCAWG dataset described in Table~\ref{tab:the_studies} using BaP Multi-NMF. Following the results of our simulation study, we  employ the recovery-discovery prior, using version 3.4 COSMIC signatures~\citep{tate_cosmic_2019} as $\mathbf{P}^{Recov}$ with $R_{recov}=86$ recoverable signatures, and allowing up to  $R_{discov}=5$ discovered signatures. We filter the $R=R_{recov}+R_{discov}=91$ estimated signatures based on $\{a^*\}$, selecting signatures whose 95th percentile of $\{a^*\}$ across all subjects from all seven cancer groups is 0 or greater. When we apply this filter, we find 43 recovered signatures and 2 discovered signatures. We refer to the discovered signatures as 'D1' and 'D2', and note they have a moderate cosine similarity of $0.64$ and $0.76$ with COSMIC signatures SBS29 and SBS17a, respectively. Compared to SBS29, D1 has a greater proportion of C>G and T>C mutations. Meanwhile, D2 contains fewer C>A, C>T and T>G mutations compared to SBS17a. We provide a visualization of signatures D1 and D2 in comparison to their closest COSMIC counterparts in Supplementary Figure S2.

\subsection{Study-level results}

\begin{figure}[!ht]
    \centering
    \includegraphics[width=0.6\linewidth]{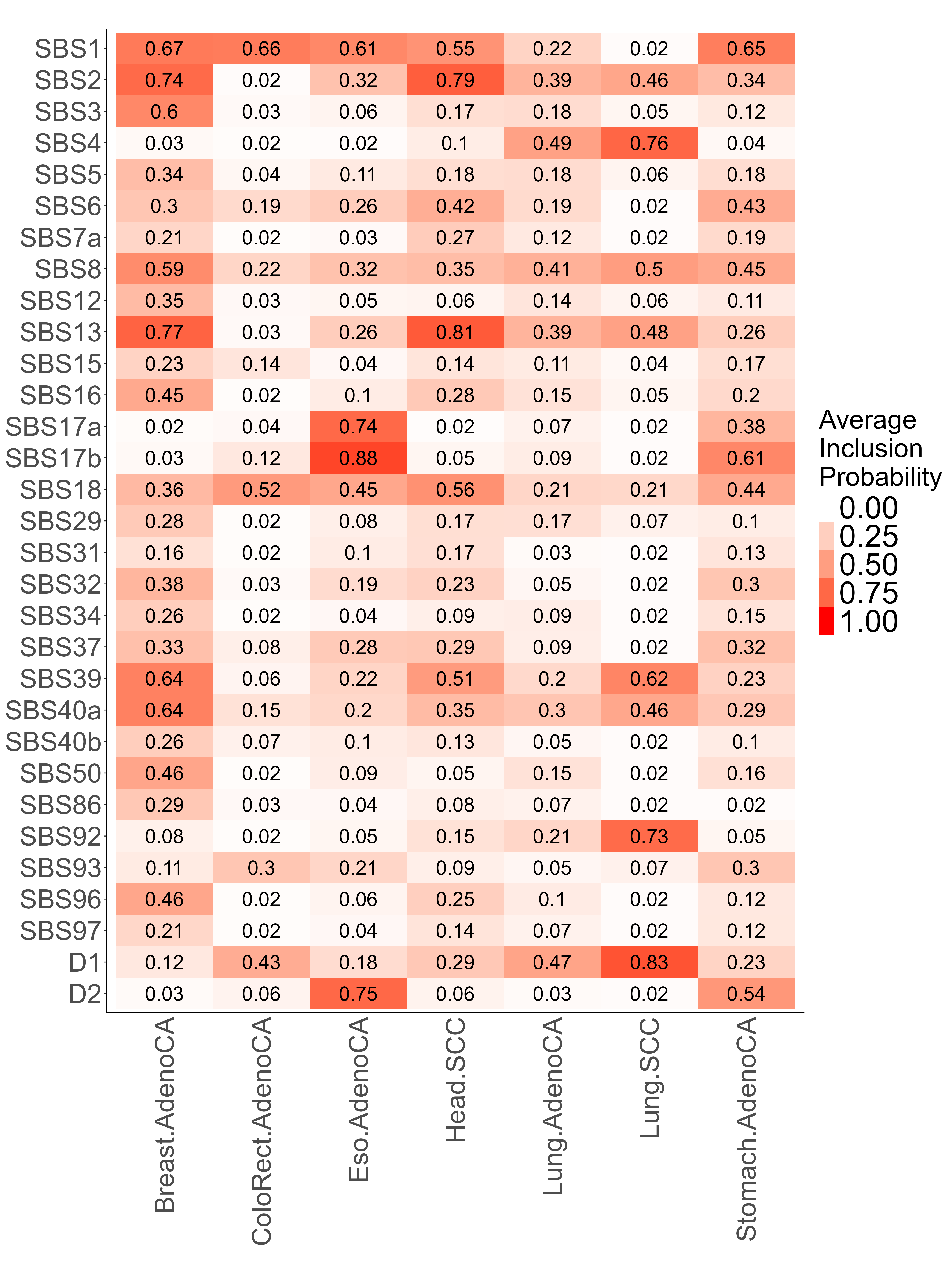}
    \caption{Average Signature Inclusion Probabilities by cancer type, calculated as $1/N_s\sum_{j=1}^{N_s}\Phi\left(\boldsymbol{\beta}_{sr}^\intercal\boldsymbol{x}_{sj}\right)$, from BaP Multi-NMF with recovery-discovery prior.}
    \label{fig:sig_inc_plot}
\end{figure}

First, we present study-level inclusion prevalence estimates for the 43 detected mutational signatures, as shown in Figure~\ref{fig:sig_inc_plot}. We find that several mutational signatures, including SBS1, SBS2, SBS8, SBS13, SBS18, and SBS39,  exhibit high prevalence in multiple cancer types, consistent with previous findings~\cite{alexandrov_repertoire_2020}. Additionally, we detect the recently defined SBS40a and SBS40b~\citep{senkin_geographic_2024}. Notably, mutational signatures associated with tobacco smoking, such as SBS4 and SBS92, are more prevalent in lung squamous cell carcinoma, which is known to be strongly associated with smoking, compared to lung adenocarcinoma and head squamous cell carcinoma, which are associated with smoking but can occur in nonsmokers~\citep{subramanian_lung_2007, dahlstrom_squamous_2008}. Moreover, we observe a comparatively higher prevalence of SBS17a and SBS17b in colorectal and stomach cancers, consistent with previous studies~\citep{secrier_mutational_2016}. Additionally, we detect many mutational signatures with high prevalence in breast cancer profiles relative to other cancer types. These include SBS3, which is known to be associated with BRCA1 and BRCA2 mutations~\citep{nik-zainal_mutational_2012}. Lastly, we detect a smaller overall number of mutational signatures in colorectal cancer. This could mean that the majority of mutational profiles from colorectal cancers are dominated by a combination of fewer signatures. 
Beyond previously identified signatures, we analyze discovered signatures D1 and D2. D1 is highly prevalent in lung squamous cell carcinoma with a lower presence in the remaining cancer types, suggesting it may represent a smoking- or tobacco-related mutational signature not captured within the current COSMIC library. The fact that this signature is found to have a relatively high prevalence in other types makes it less likely that it is an artifact associated with a single study. D2 is present in the profiles of colorectal adenocarcinoma and stomach adenocarcinoma and appears to be correlated with SBS17a and SBS17b. Given the small sample size in this analysis, further validation of D1 and D2 is necessary. 

\subsection{Subject-level results}

\begin{figure}[!ht]
    \centering
    \includegraphics[width=\linewidth]{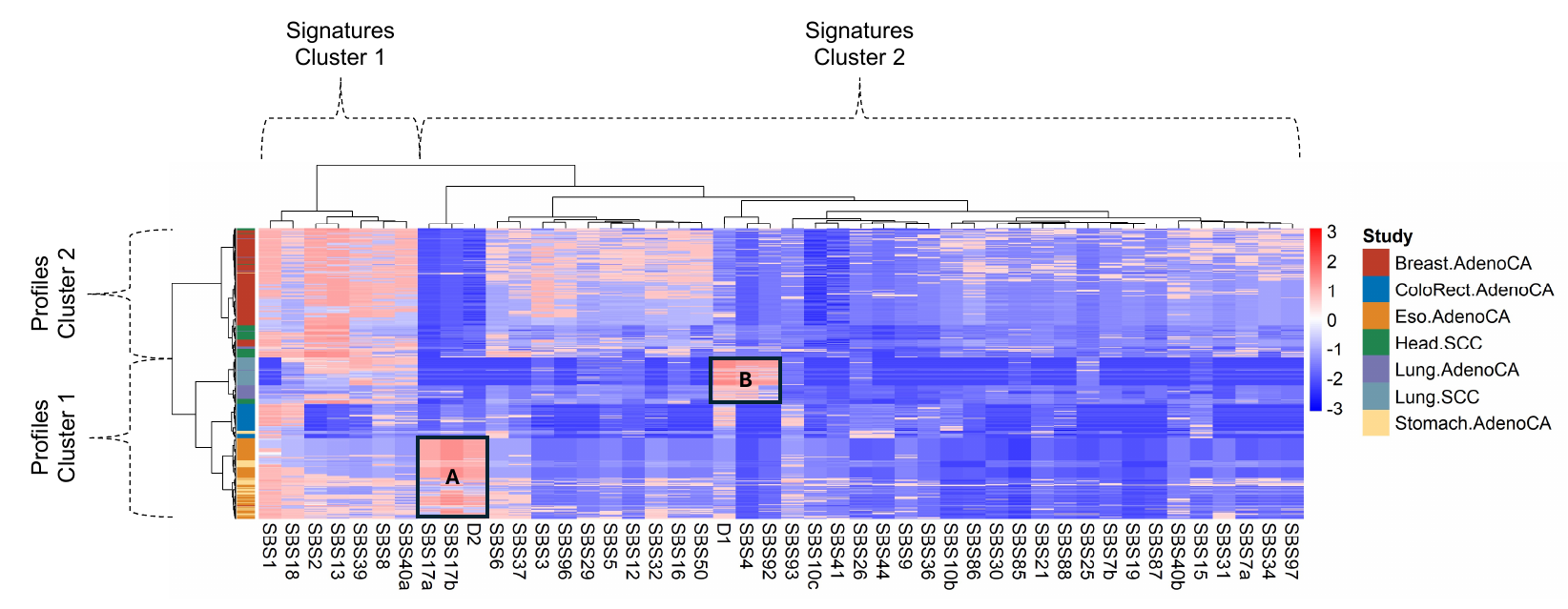}
    \caption{Heatmap visualizing subject-level inclusion probabilities of mutational signatures across all samples in the 7 cancer types considered. In this matrix, each row corresponds to a mutational profile, each column corresponds to a mutational signature, and the color of the heatmap is determined by the posterior mean of $a^*_{sjr}$. Mutational profiles and mutational signatures are both clustered using complete-linkage clustering using Euclidean distance as the metric. The left column denotes the cancer type of the mutational profile, and is not directly used in clustering. Box A highlights exposures to clustered signatures SBS17a, SBS17b, and D2 within colorectal and stomach cancer profiles, and Box B highlights exposures to clustered signatures D1, SBS4, and SBS92 among lung cancer profiles.}
    \label{fig:pheatmap-fig}
\end{figure}

Next, we analyze the information provided by the subject-level signature inclusion indicators, $a^*_{sjr}$, a unique feature of our modeling approach. Figure~\ref{fig:pheatmap-fig} visualizes $a^*_{sjr}$ in all subjects in the seven studies, arranged via complete-linkage clustering of both the subjects (rows) and mutational signatures (columns) using the Euclidean distance metric. This analysis reveals structure of both mutational signatures and cancer mutational profiles. 
Among the mutational signatures, we find two main clusters. Signature cluster 1 includes mutational signatures,  such as SBS1, SBS2, SBS8, SBS18, SBS39, and SBS40a,  which are present within many samples. Signatures cluster 2 contains the remaining 37 mutational signatures, which are detected in a subset of the samples. Within signatures cluster 2, we observe several sub-clusters, including one containing SBS17a, SBS17b, and D2 (Box A), and another containing SBS4, SBS92, and D1 (Box B). These sub-clusters align with  previous findings, as SBS17a and SBS17b often co-occur in esophagus and stomach cancers, while SBS4, SBS92 and SBS29 are associated with tobacco smoking and chewing~\citep{alexandrov_signatures_2013, alexandrov_clock-like_2015, secrier_mutational_2016, lawson_extensive_2020}.

\begin{figure}
    \centering
    \includegraphics[width=\linewidth]{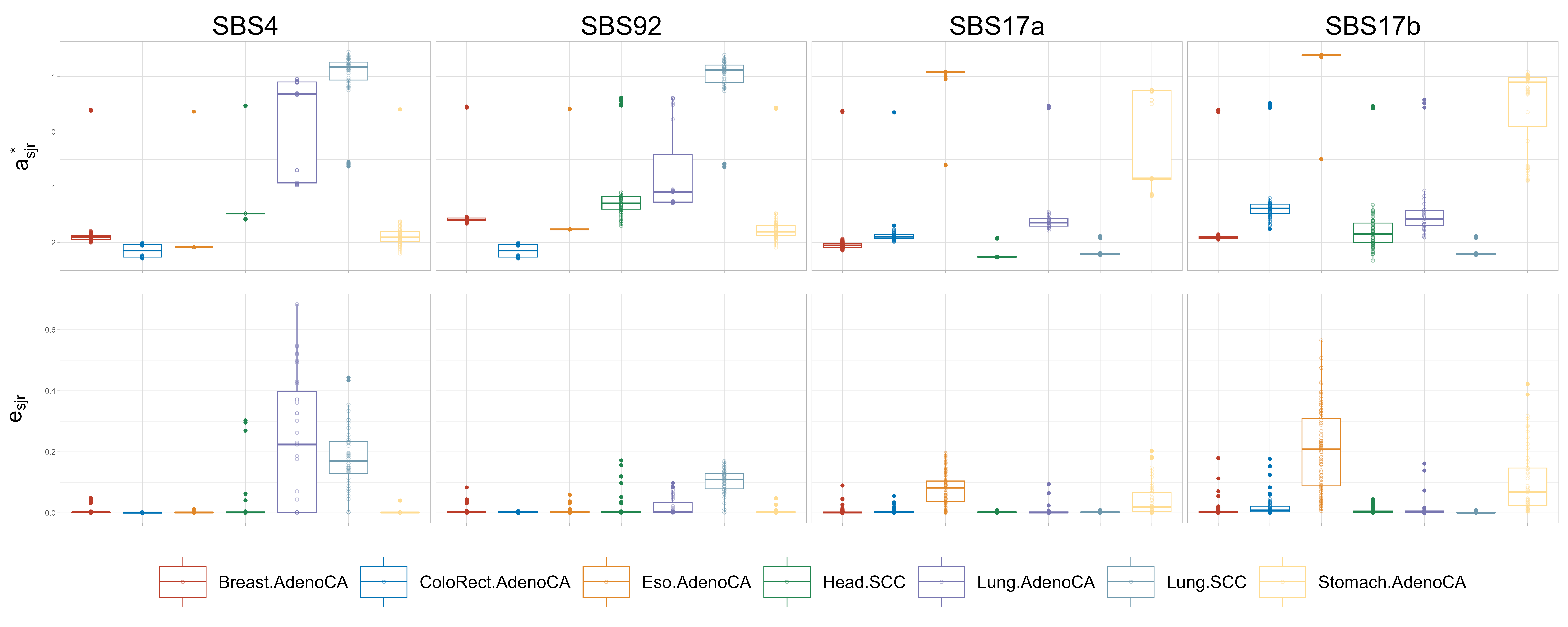}
    \caption{Distribution of latent inclusion parameter, $a^*_{sjr}$, and the proportion of mutations, $e_{sjr}$, in selected mutational signatures across the 7 cancer types.}
    \label{fig:panels}
\end{figure}

Individual profiles can be roughly divided, to begin, into two primary groups. Profile cluster 2 includes a sub-cluster of both lung cancer types and some head cancers, and another sub-cluster containing colorectal, stomach, and esophagus cancers. Profile cluster 1 consists of mutational profiles from breast and head cancers, along with a few lung adenocarcinomas. An interesting finding is that the head cancers that cluster with the lung samples in Cluster 2 appear to have higher $a^*_{sjr}$ among the smoking-related signatures compared to the other head cancers that cluster more closely to the breast cancer samples. An opposite trend is present with the lung cancers that cluster with the head cancers in Cluster 1 - they have relatively lower $a^*_{sjr}$ in the smoking related cancers. These results are reflected in Figure~\ref{fig:panels}, which visualizes $a^*_{sjr}$ and $e_{sjr}$ in smoking-related signatures SBS4 and SBS92 across the seven cancer types. We find a small subset of head squamous cell carcinoma profiles with $e_{sjr}>0.1$ for SBS4 and SBS92, which could explain why they cluster more closely with the lung cancer profiles. We can also see in Figure~\ref{fig:panels} that SBS17a/SBS17b were relevant in a meaningful proportion of the stomach and esophagus cancers, with a small number of colorectal cancers having some exposure to SBS17b.

\begin{figure}
    \centering
    \includegraphics[width=0.5\linewidth]{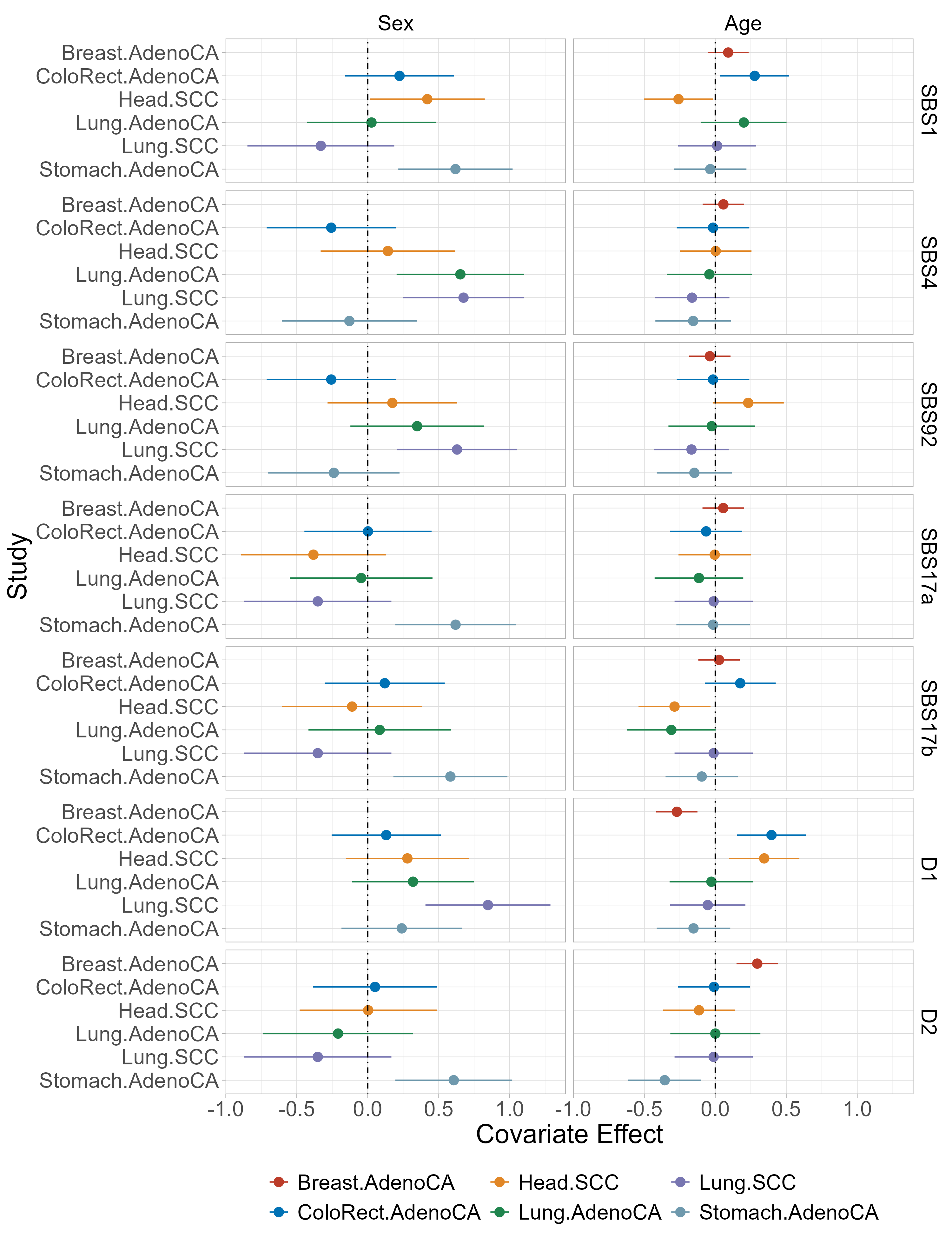}
    \caption{Visualization of study-specific covariate effects, reported as a 95\% credible interval based on the variational approximation of BaP Multi-NMF. Due to limitations noted in Section~2, we excluded the sex covariate for the breast adenocarcinoma samples, and we excluded both covariates for esophagus adenocarcinoma.}
    \label{fig:pcawg_covariates}
\end{figure}

In addition to analyzing subject-level mutational signature presence, we also explore study-specific covariate effects that influence the presence of mutational signatures, another unique feature of our model. Figure~\ref{fig:pcawg_covariates} visualizes the estimated covariate effects for several mutational signatures (complete results are visualized in Supplementary Figure S3). We found a positive association between the presence of SBS1 and age in 3 out of 6 types of cancer with sufficient age data. This finding aligns with previous studies showing that SBS1 is a clock-like signature associated with mutations that accumulate over time~\citep{nik-zainal_mutational_2012}. We found positive associations between smoking mutational signatures SBS4 and SBS92 and male sex in lung adenocarcinoma and lung squamous cell carcinoma, consistent with the known higher prevalence of tobacco use among men~\citep{israel_t_tobacco_2014}. Furthermore, we found that SBS17a and SBS17b were associated with male sex in the stomach adenocarcinoma group. Among the signatures discovered, D1 was associated with male sex in lung squamous cell carcinoma with different age effects in breast, colorectal and head cancer. 
Signature D2 was predominantly associated with younger and male subjects in the stomach adenocarcinoma group. Future work is needed to validate the discovered signatures D1 and D2 and associate them with a mutagenic process.

\section{Discussion}

In this paper, we presented a new Bayesian Multi-Study NMF model capable of analyzing multi-study mutational signature data with covariates to simultaneously estimate mutational signatures and detect their presence at the subject level. Our implementation leverages variational Bayes, which offers computationally efficiency and accuracy, particularly in scenarios with a large number of mutational profiles. Additionally, we provided a powerful option to incorporate previously discovered mutational signatures through a recovery-discovery prior.

We note several remaining limitations of BaP Multi-NMF. As highlighted in Section~\ref{sec:sims}, the model may fail to detect mutational signatures that comprise a small fraction of mutations in a particular mutational profile. Although their biological significance may be lower compared to signatures that explain a greater proportion of mutations within mutational profiles, addressing this limitation could further enhance the model's utility. Future directions could including developing a version of BaP Multi-NMF that models exposures as a mixture of gamma random variables, allowing for the detection of mutational signatures with low exposures.

Additionally, there remain several practical and statistical complications inherent to NMF-based approaches. Without imposing  additional constraints, issues of identifiability persist for both the mutational signatures and exposures matrix. A consequence is that any optimization procedure, such as optimizing the ELBO with a variational approach, is highly non-convex and sensitive to initialization. Similarly, standard Bayesian methods can also struggle with multi-modality present within the posterior distribution, and a successful MCMC implementation of BaP Multi-NMF would likely require a complex sampling scheme such as the tempered sampler used by \cite{grabski_bayesian_2023}. Alternatively, incorporating a repulsive prior on the mutational signature matrix, which discourages proximity between signatures based on a defined distance metric~\citep{petralia_repulsive_2012}, could provide an elegant solution to these challenges.


\begin{funding}
B.H. was supported by the US National Institutes of Health, under grants NIGMS/NIH COBRE CBHD P20GM109035and. R.D.V. was supported by the US National Institutes of Health, under grants NIGMS/NIH COBRE CBHD P20GM109035, and 5R01CA262710. G.P. was supported by the U.S.A. National Institutes of Health, through grants 5R01CA262710 and 5R01CA240299 and by the U.S.A. National Science Foundation through grant DMS 2113707.  I.N.G. is the Kenneth G. Langone Quantitative Biology Fellow of the Damon Runyon Cancer Research Foundation (DRQ-21-24).
\end{funding}




\bibliographystyle{imsart-nameyear} 
\bibliography{refs}       

\end{document}


\maketitle
\section{Supporting Calculations}\label{sec:supporting_calcs}
In this Section, we derive conditional posterior distributions for each component of the mean field and use those results to derive CAVI updates. 

In order to estimate the parameters $\boldsymbol{\theta}$ of the BAP Multi-NMF model, we approximate the posterior distribution $p(\boldsymbol{\theta} \mid \left\{\mathbf{M}_s\right\})$ via a class of densities that are members of the mean-field:
\begin{align}
    q(\boldsymbol{\theta}) &= \left[\prod_{r=1}^R q(\mathbf{P}_r)\right] \left[\prod_{s=1}^S\prod_{j=1}^{N_s}q\left(\mathbf{E}_{sj}\right)\right] \left[\prod_{s=1}^S\prod_{j=1}^{N_s}q(w_{sj})\right] \\
    &\times \left[ \prod_{s=1}^S\prod_{j=1}^{N_s}\prod_{r=1}^R q(a_{sjr})q(a^*_{sjr})\right]\left[ \prod_{s=1}^S\prod_{r=1}^R q(\boldsymbol{\beta}_{sr})q(\tau_{sr}) \right]  \left[ \prod_{s=1}^S q(\alpha_s^w) q(\beta_s^w)  \right], 
\end{align}
where we refer to each component of the mean field a variational factor, or factor for short. The optimal approximation, $q(\boldsymbol{\theta}^*)$, satisfies the minimization:
\begin{equation}\label{ELBO}
    \underset{q(\boldsymbol{\theta}) \in \mathcal{Q}}{\arg\max}\left\{ \text{ELBO}(q(\boldsymbol{\theta}))\right\}= \underset{q(\boldsymbol{\theta}) \in \mathcal{Q}}{\arg\max}\left\{ \mathbb{E}_{q(\boldsymbol{\theta})}[\log p( \boldsymbol{\theta} , \mathbf{X} )] - \mathbb{E}_{q(\boldsymbol{\theta})}[\log q(\boldsymbol{\theta})]\right\}.
\end{equation}
We estimate $q(\boldsymbol{\theta}^*)$ via coordinate-ascent variational inference (CAVI, \citep{bishop_pattern_2006}), using the following equation to update factor $m$ of $q(\boldsymbol{\theta})$ conditional on the remaining $m-1$ factors:
\begin{equation}\label{cavi-update}
    q^*_m(\boldsymbol{\theta}_m) \propto \exp \left( \mathbb{E}_{q(\boldsymbol \theta_{-m})}\left[\log(p(\boldsymbol{\theta}_m | \boldsymbol{\theta}_{-m}, \left\{\mathbf{M}_s\right\})\right]\right).
\end{equation}

\subsection{Conditional Posterior Derivations and CAVI Updates}\label{sec:cavi_updates}

\subsubsection{Latent Mutation Counts}
A challenge of NMF models is that many parameter updates depend on signature specific mutation counts, $z_{sijr}$, which is not observed. Instead we observe $M_{sij}=\sum_{r=1}^R z_{sijr}$. Assuming they are observed, we can use the complete data likelihood:
\begin{equation}
    l\left(\boldsymbol{\theta} \mid \left\{\mathbf{M}_s\right\}, \left\{z_{sijr}\right\}\right) = \prod_{s=1}^S\prod_{i=1}^K\prod_{r=1}^R\prod_{j=1}^{N_s} \left(w_{sj}p_{ir}e_{sjr}\right)^{z_{sijr}}\left(z_{sijr}\right)^{-1}  \mathbbm{1} {\left\{\sum_{r} z_{sijr}=m_{sij}\right\}}.
\end{equation}
Note have $z_{sijr} \sim \text{Poisson}\left(w_{sj}p_{ir}e_{sjr}\right)$ and $\sum_{r=1}^R z_{sijr} = m_{sij}$. Therefore,
\begin{align}
    z_{sij1},\ldots,z_{sijR} \mid \ldots &\sim \text{Multinomial}\left(m_{sij}, \mathbf{p}_{sij}^z\right),
\end{align}
where:
\begin{equation}
    p_{sijr}^z=\frac{w_{sj}p_{ir}e_{sjr}}{\sum_{r}w_{sj}p_{ir}e_{sjr}}=\frac{p_{ir}e_{sjr}}{\sum_{r}p_{ir}e_{sjr}}.
\end{equation}
For the purposes of performing CAVI, we set $z_{sijr}=\mathbb{E}\left[ z_{sijr}\right]$ at each iteration and use the complete data likelihood for the remaining computations.

\subsubsection{Signatures}
The conditional posterior distribution for $p_{ir}$ is given by:
\begin{align}
    p(p_{ir}\mid\ldots) &\propto \left(\prod_{s=1}^S\prod_{j=1}^{N_s} [p_{ir}]^{z_{sijr}}\right)[p_{ir}]^{\alpha_{ir}^p -1}, \\
    &= [p_{ir}]^{\alpha_{ir}^p + \sum_s\sum_j z_{sijr} - 1},
\end{align}
which implies: 
\begin{align}
    p_{1r},\ldots,p_{Kr} \mid \ldots &\sim \text{Dirichlet}\left(a_{1r}^p,\ldots,a_{Kr}^p\right),
\end{align}
where $a_{ir}^p = \alpha_{ir}^p + \sum_s \sum_j z_{sijr}$. The optimal factor for mutational signatures parameter $\mathbf{p}_{r}$ is given by:
\begin{align}
    \log q^*\left(\mathbf{p}_{r}\right) &\propto  \mathbb{E}_{q(\boldsymbol \theta_{-\mathbf{p}_{r}})}\left[\log(p(\mathbf{p}_{r} | \dots )\right], \\
    &\propto \mathbb{E}_{q(\boldsymbol \theta_{-\mathbf{p}_{kr}})}\left[\sum_{k=1}^K \left(\alpha_{ir}^p + \sum_s\sum_j z_{sijr} - 1\right) p_{kr}\right], \\
    &= \sum_{k=1}^K \left(\alpha_{ir}^p + \sum_s\sum_j z_{sijr} - 1\right) p_{kr},
\end{align}
where $\mathbf{p}_{r}=p_{1r},\dots,p_{Kr}$. This implies $q^*\left(\mathbf{p}_{kr}\right)=\text{Dirichlet}\left(\theta_{1r}^p,\ldots,\theta_{Kr}^p\right)$, where
\begin{equation}
    \theta_{ir}^p = \alpha_{ir}^p + \sum_s \sum_j z_{sijr},
\end{equation}
with notation $\boldsymbol{\theta}_{r}^p=\theta_{1r}^p,\dots, \theta_{Kr}^p.$

\subsubsection{Exposures}
The conditional posterior distribution for $e_{srj}$ is given by:
\begin{align}
    p(e_{srj}\mid\dots) &\propto \left(\prod_{k=1}^K \left(e_{sjr}\right)^{z_{sijr}} \right) \left(e_{sjr}\right)^{(a_{sjr})\alpha_s^{e1} + (1-a_{sjr})\alpha_s^{e0} - 1}, \\
    &= \left(e_{sjr}\right)^{\sum_{k=1}^K z_{sijr} + (a_{sjr})\alpha_s^{e1} + (1-a_{sjr})\alpha_s^{e0} - 1},
\end{align}
which implies:
\begin{align}
    e_{s1j},\ldots,e_{sRj} \mid \ldots &\sim \text{Dirichlet}\left(a_{s1j}^e,\ldots,a_{sRj}^e\right),
\end{align}
where $a_{srj}^e = (a_{sjr})\alpha_s^{e1} + (1-a_{sjr})\alpha_s^{e0} + \sum_{k=1}^K z_{sijr}$. The optimal factor for exposure factor $e_{srj}$ is given by:
\begin{align}
 \log q^*\left(\mathbf{e}_{sj}\right) &\propto  \mathbb{E}_{q(\boldsymbol \theta_{-\mathbf{e}_{sj}})}\left[\log(p( \mathbf{e}_{sj} | \dots )\right], \\
 &\propto \mathbb{E}_{q(\boldsymbol \theta_{-e_{srj}})}\left[ \sum_{r=1}^R \left((a_{sjr})\alpha_s^{e1} + (1-a_{sjr})\alpha_s^{e0} + \sum_{k=1}^K z_{sijr}\right) e_{srj}  \right], \\
 &= \sum_{r=1}^R \left( \mathbb{E}\left[a_{sjr}\right]\alpha_s^{e1} + (1- \mathbb{E}\left[a_{sjr}\right])\alpha_s^{e0} + \sum_{k=1}^K z_{sijr}\right) e_{srj},
\end{align}
where $\mathbf{e}_{sj} = e_{s1j},\dots,e_{sRj}$. This implies $q^*\left(\mathbf{e}_{sj}\right)=\text{Dirichlet}\left(\theta_{s1j}^e,\dots, \theta_{sRj}^e\right)$, where
\begin{equation}
\theta_{srj}^e=\mathbb{E}\left[a_{sjr}\right]\alpha_s^{e1} + (1- \mathbb{E}\left[a_{sjr}\right])\alpha_s^{e0} + \sum_{k=1}^K z_{sijr},
\end{equation}
with $\boldsymbol{\theta}_{sj}^e = \theta_{s1j}^e,\dots,\theta_{sRj}^e$.

Next, we can infer:
\begin{align}
    p(a_{srj}\mid\dots = 1) &\propto p\left(e_{srj}\mid a_{srj} =1 \right)p\left(a_{srj} =1\right),\\
    &= \frac{\Gamma\left(\alpha_s^{e1} + \sum_{t \neq r} a_{stj} \alpha_s^{e1} + (1-a_{stj}) \alpha_s^{e0}  \right)}{\Gamma\left(\alpha_s^{e1}\right)\prod_{t \neq r} \Gamma(a_{stj} \alpha_s^{e1} + (1-a_{stj}) \alpha_s^{e0} )}\left(e_{srj}\right)^{\alpha_s^{e1}-1}\left(1-\Phi\left(-\boldsymbol{\beta}_{sr}^\intercal\boldsymbol{x}_{sj}\right)\right),\\
    p(a_{srj}\mid\dots = 0) &\propto  p\left(e_{srj}\mid a_{srj} =0 \right)p\left(a_{srj} =0\right),\\
    &= \frac{\Gamma\left(\alpha_s^{e0} + \sum_{t \neq r} a_{stj} \alpha_s^{e1} + (1-a_{stj}) \alpha_s^{e0}  \right)}{\Gamma\left(\alpha_s^{e0}\right)\prod_{t \neq r} \Gamma(a_{stj} \alpha_s^{e1} + (1-a_{stj}) \alpha_s^{e0} )}\left(e_{srj}\right)^{\alpha_s^{e0}-1}\Phi\left(-\boldsymbol{\beta}_{sr}^\intercal\boldsymbol{x}_{sj}\right),
\end{align}
which implies $p(a_{srj}\mid \dots) = \text{Bernoulli}\left(\pi_{srj}^a\right)$, where:
\begin{equation}
    \pi_{srj}^a=\frac{p(a_{srj}\mid\dots = 1)}{p(a_{srj}\mid\dots = 1) + p(a_{srj}\mid\dots = 0)}.
\end{equation}
The optimal factor for $a_{srj}$ is given by:
\begin{align}
 \log q^*\left(a_{srj}\right) &\propto  \mathbb{E}_{q(\boldsymbol \theta_{-a_{srj}})}\left[\log(p( a_{srj} | \dots )\right], \\
 &\propto \mathbb{E}_{q(\boldsymbol \theta_{-a_{srj}})}\left[ a_{srj} \log \pi_{srj}^a + \left(1 - a_{srj}\right) \log \left(1 - \pi_{srj}^a\right) \right] \\
 &\propto a_{srj}  \mathbb{E}_{q(\boldsymbol \theta_{-a_{srj}})}\left[ \log \frac{\pi_{srj}^a}{1 - \pi_{srj}^a}\right],
\end{align}
which implies $ q^*\left(a_{srj}\right) = \text{Bernoulli}\left(\theta_{srj}^a\right)$, where 
\begin{align}
    \theta_{srj}^a &= \frac{1}{1+\exp{-\mathbb{E}_{q(\boldsymbol \theta_{-a_{srj}})}\left[ \log \frac{\pi_{srj}^a}{1 - \pi_{srj}^a}\right]}}, \\
    \mathbb{E}_{q(\boldsymbol \theta_{-a_{srj}})}\left[ \log \frac{\pi_{srj}^a}{1 - \pi_{srj}^a}\right] &=  \mathbb{E}_{q(\boldsymbol \theta_{-a_{srj}})}\left[\log\frac{p(a_{srj}\mid\dots = 1)}{p(a_{srj}\mid\dots = 0)}\right], \\
    &= \mathbb{E}_{q(\boldsymbol \theta_{-a_{srj}})}\left[\log p(a_{srj}\mid\dots = 1) - \log p(a_{srj}\mid\dots = 0)\right], \\
    &= \mathbb{E}_{q(\boldsymbol \theta_{-a_{srj}})}\left[\log\Gamma\left(\alpha_s^{e1} + \sum_{t \neq r} a_{stj} \alpha_s^{e1} + (1-a_{stj}) \alpha_s^{e0}  \right) \right. \\
    &\qquad \left. -\log \Gamma\left(\alpha_s^{e1}\right) + \left(\alpha_s^{e1} - 1\right) \log e_{srj} + \log \left(1-\Phi\left(-\boldsymbol{\beta}_{sr}^\intercal\boldsymbol{x}_{sj}\right)\right)  \right. \\
    &\qquad \left. -  \log\Gamma\left(\alpha_s^{e0} + \sum_{t \neq r} a_{stj} \alpha_s^{e1} + (1-a_{stj}) \alpha_s^{e0}  \right) \right. \\
    &\qquad \left. +\log \Gamma\left(\alpha_s^{e0}\right) - \left(\alpha_s^{e0} -1\right)\log e_{srj} - \log \left(\Phi\left(-\boldsymbol{\beta}_{sr}^\intercal\boldsymbol{x}_{sj}\right)\right)\right], \\
    &= \mathbb{E}_{q(\boldsymbol \theta_{-a_{srj}})}\left[\log\Gamma\left(\alpha_s^{e1} + \sum_{t \neq r} a_{stj} \alpha_s^{e1} + (1-a_{stj}) \alpha_s^{e0}  \right) \right] \\
    &\qquad  - \mathbb{E}_{q(\boldsymbol \theta_{-a_{srj}})}\left[\log \Gamma\left(\alpha_s^{e1}\right)\right] +  \left(\mathbb{E}_{q(\boldsymbol \theta_{-a_{srj}})}\left[\alpha_s^{e1}\right] - 1\right) \log e_{srj} \\
    &\qquad +  \mathbb{E}_{q(\boldsymbol \theta_{-a_{srj}})}\left[\log \left(1-\Phi\left(-\boldsymbol{\beta}_{sr}^\intercal\boldsymbol{x}_{sj}\right)\right)\right]   \\
    &\qquad  -   \mathbb{E}_{q(\boldsymbol \theta_{-a_{srj}})}\left[\log\Gamma\left(\alpha_s^{e0} + \sum_{t \neq r} a_{srj} \alpha_s^{e1} + (1-a_{stj}) \alpha_s^{e0}  \right)\right]\\
    &\qquad + \mathbb{E}_{q(\boldsymbol \theta_{-a_{srj}})}\left[\log \Gamma\left(\alpha_s^{e0}\right)\right] -  \left(\mathbb{E}_{q(\boldsymbol \theta_{-a_{srj}})}\left[\alpha_s^{e0}\right] -1 \right) \log e_{srj} \\
    &\qquad -  \mathbb{E}_{q(\boldsymbol \theta_{-a_{srj}})}\left[\log \left(\Phi\left(-\boldsymbol{\beta}_{sr}^\intercal\boldsymbol{x}_{sj}\right)\right)\right].
\end{align}
In our implementation of CAVI, we simplify the above by using the approximation \citep{bowling_logistic_2009}:
$$
\Phi(x) \approx \left(1 + \exp(-\left(0.7056x^3 + 1.5976 x \right))\right)^{-1}.
$$
Furthermore, several expectations $\mathbb{E}\left[f\left(x\right)\right]$ are approximated by a second-order taylor series expansion of $f$ centered at $\mathbb{E}\left[x\right]$, a step which has been taken in other variational approaches~\citep{ishiguro_averaged_2017}:
\begin{align*}
    \mathbb{E}\left[f\left(x\right)\right] &\approx f\left(\mathbb{E}\left[x\right]\right) + \frac{1}{2}\mathbb{V}(x) f^{''}\left(\mathbb{E}\left[x\right]\right).
\end{align*}

Moving on, the conditional posterior for $a^*_{sjr}$ is given by:
\begin{align}
    p(a^*_{sjr}\mid\dots) &\propto \begin{cases}
        \mathcal{N}_{[0, \infty)}\left(\boldsymbol{\beta}_{sr}^\intercal\boldsymbol{x}_{sj},1\right) & a_{sjr} = 1 \\
        \mathcal{N}_{(-\infty, 0]}\left(\boldsymbol{\beta}_{sr}^\intercal\boldsymbol{x}_{sj},1\right) & a_{sjr} = 0
    \end{cases}, \\
    &= \mathcal{N}_{[0, \infty)}\left(\boldsymbol{\beta}_{sr}^\intercal\boldsymbol{x}_{sj},1\right)^{a_{sjr}}\mathcal{N}_{(-\infty, 0]}\left(\boldsymbol{\beta}_{sr}^\intercal\boldsymbol{x}_{sj},1\right)^{1-a_{sjr}},
\end{align}
where $\mathcal{N}_{[0, \infty)}\left(\cdot\right)$, $\mathcal{N}_{(-\infty, 0]}\left(\cdot\right)$ refers to the normal distributed truncated above or below zero, respectively. The optimal factor for $a^*_{srj}$ is given by:
\begin{align}
 \log q^*\left(a^*_{srj}\right) &\propto  \mathbb{E}_{q(\boldsymbol \theta_{-a^*_{srj}})}\left[\log(p( a^*_{srj} | \dots )\right], \\
 &\propto \mathbb{E}_{q(\boldsymbol \theta_{-a^*_{srj}})}\left[-a_{sjr} \mathbbm{1}\left\{a^*_{srj} > 0\right\}\left(a^*_{srj}-\boldsymbol{\beta}_{sr}^\intercal\boldsymbol{x}_{sj}\right)^2 - \left(1-a_{sjr} \right) \mathbbm{1}\left\{a^*_{srj} < 0\right\}\left(a^*_{srj}-\boldsymbol{\beta}_{sr}^\intercal\boldsymbol{x}_{sj}\right)^2\right], \\
 &\propto -\mathbb{E}_{q(\boldsymbol \theta_{-a^*_{srj}})}\left[a_{sjr}\right] \mathbbm{1}\left\{a^*_{srj} > 0\right\}\left(a^*_{srj}-\mathbb{E}_{q(\boldsymbol \theta_{-a^*_{srj}})}\left[\boldsymbol{\beta}_{sr}\right]^\intercal\boldsymbol{x}_{sj}\right)^2  \label{a_star_calc_expand_1} \\
 &\qquad  - \left(1-\mathbb{E}_{q(\boldsymbol \theta_{-a^*_{srj}})}\left[a_{sjr}\right] \right) \mathbbm{1}\left\{a^*_{srj} < 0\right\}\left(a^*_{srj}- \mathbb{E}_{q(\boldsymbol \theta_{-a^*_{srj}})}\left[\boldsymbol{\beta}_{sr}\right]^\intercal\boldsymbol{x}_{sj}\right)^2\label{a_star_calc_expand_2}, \\
 q^*\left(a^*_{srj}\right) &= p\left(a_{srj}=1\right)\mathcal{N}_{[0, \infty)}\left(\theta_{srj}^{a*},1\right) + p\left(a_{srj}=0\right)\mathcal{N}_{(-\infty, 0]}\left(\theta_{srj}^{a*},1\right)
\end{align}
where $\theta_{srj}^{a*}=\left[\boldsymbol{\beta}_{sr}\right]^\intercal\boldsymbol{x}_{sj}$ and lines \eqref{a_star_calc_expand_1}-\eqref{a_star_calc_expand_2} follow by expanding the quadratic term, taking expectations, and completing the square. 

The conditional posterior for covariate effect vector $\boldsymbol{\beta}_{sr}$ is given by:
\begin{align}
    p(\boldsymbol{\beta}_{sr} \mid \dots) &\propto \mathcal{MVN}\left(\mathbf{A}^*_{sr}\mid \mathbf{X}_s\boldsymbol{\beta}_{sr} \mathcal{I}_{N_s}\right)\mathcal{MVN}\left(\boldsymbol{\beta}_{sr}\mid(\boldsymbol{\beta}^0_{sr}, \tau_{sr}^{-1}\mathcal{I}_{Q_s})\right), \\
    &= \mathcal{MVN}\left(\boldsymbol{\beta}_{sr} \mid \boldsymbol{\mu}_{sr}^{\beta}, \boldsymbol{\Sigma}_{sr}^{\beta} \right)
\end{align}
where $\mathbf{A}^*_{sr} = \left(a^*_{s1r},\ldots,a^*_{sN_sr}\right)^\intercal$, $\mathbf{X}_s = \left(\boldsymbol{x}_{s1}, \ldots, \boldsymbol{x}_{sN_s}\right)^\intercal$, $Q_s$ is the number of covariates in study $s$ and
\begin{align}
    \boldsymbol{\Sigma}_{sr}^{\beta} &= \left(\tau_{sr}\mathcal{I}_{Q_s} + \mathbf{X}_s^\intercal \mathbf{X}_s\right)^{-1}, \\
    \boldsymbol{\mu}_{sr}^{\beta} &= \boldsymbol{\Sigma}_{sr}^{\beta}\left(\tau_{sr}\mathcal{I}_{Q_s}\boldsymbol{\beta}^0_{sr} + \mathbf{X}_s^\intercal\mathbf{A}^*_{sr}\right).
\end{align}
The optimal factor for $\boldsymbol{\beta}_{sr}$ is given by:
\begin{align}
 \log q^*\left(\boldsymbol{\beta}_{sr}\right) &\propto  \mathbb{E}_{q(\boldsymbol \theta_{-\boldsymbol{\beta}_{sr}})}\left[\log(p( \boldsymbol{\beta}_{sr} | \dots )\right], \\
 &\propto \mathbb{E}_{q(\boldsymbol \theta_{-\boldsymbol{\beta}_{sr}})}\left[-\frac{1}{2}\left(\boldsymbol{\beta}_{sr}-\boldsymbol{\mu}_{sr}^{\beta}\right)^\intercal\left[\boldsymbol{\Sigma}_{sr}^{\beta}\right]^{-1}\left(\boldsymbol{\beta}_{sr}-\boldsymbol{\mu}_{sr}^{\beta}\right)\right], 
\end{align}
which implies 
\begin{equation}
    q^*\left(\boldsymbol{\beta}_{sr}\right) = \mathcal{MVN}\left( \boldsymbol{\theta}_{sr}^{\beta\mu}, \boldsymbol{\theta}_{sr}^{\beta\Sigma} \right),
\end{equation}
where
\begin{align}
    \boldsymbol{\theta}_{sr}^{\beta\Sigma} &= \left(\mathbb{E}\left[\tau_{sr}\right]\mathcal{I}_{Q_s} + \mathbf{X}_s^\intercal \mathbf{X}_s\right)^{-1}, \\
    \boldsymbol{\theta}_{sr}^{\beta\mu} &= \boldsymbol{\Sigma}_{sr}^{\beta}\left(\mathbb{E}\left[\tau_{sr}\right]\mathcal{I}_{Q_s}\boldsymbol{\beta}^0_{sr} + \mathbf{X}_s^\intercal\mathbb{E}\left[\mathbf{A}^*_{sr}\right]\right).
\end{align}

The conditional posterior for $\tau_{sr}$ is given by:
\begin{align}
    p(\tau_{sr}\mid\dots) &\propto \left(\boldsymbol{\beta}_{sr}\mid(\boldsymbol{\beta}^0_{sr}, \tau_{sr}^{-1}\mathcal{I}_{Q_s})\right) p(\tau_{sr}), \\
    &\propto \left(\tau_{sr}\right)^{Q_s/2}\exp\left\{-\tau_{sr}\left(\frac{1}{2}(\boldsymbol{\beta}_{sr} - \boldsymbol{\beta}^0_{sr})^\intercal(\boldsymbol{\beta}_{sr} - \boldsymbol{\beta}^0_{sr})\right)\right\}\left(\tau_{sr}\right)^{\gamma_1-1}\exp\left\{-\gamma_2\tau_{sr}\right\}, \\
    &= \left(\tau_{sr}\right)^{Q_s/2 + \gamma_1 - 1}\exp\left\{-\tau_{sr}\left(\gamma_2 + \frac{1}{2}(\boldsymbol{\beta}_{sr} - \boldsymbol{\beta}^0_{sr})^\intercal(\boldsymbol{\beta}_{sr} - \boldsymbol{\beta}^0_{sr})\right)\right\},
\end{align}
which implies 
\begin{equation}
    p(\tau_{sr}\mid\dots) \sim \Gamma(\alpha_{sr}^\tau, \beta_{sr}^\tau)
\end{equation}
where $\alpha_{sr}^\tau = Q_s/2 + \gamma_1$ and $ \beta_{sr}^\tau=\gamma_2 + \frac{1}{2}(\boldsymbol{\beta}_{sr} - \boldsymbol{\beta}^0_{sr})^\intercal(\boldsymbol{\beta}_{sr} - \boldsymbol{\beta}^0_{sr})$. The optimal factor for $\tau_{sr}$ is given by:
\begin{align}
    q^*\left(\tau_{sr}\right) &= \Gamma\left(\theta_{sr}^{\tau\alpha
    }, \theta_{sr}^{\tau\beta
    }\right), 
\end{align}
where
\begin{align}
    \theta_{sr}^{\tau\alpha
    } &=  Q_s/2 + \gamma_1, \\
    \theta_{sr}^{\tau\beta
    } &= \gamma_2 + \frac{1}{2}\mathbb{E}\left[(\boldsymbol{\beta}_{sr} - \boldsymbol{\beta}^0_{sr})^\intercal(\boldsymbol{\beta}_{sr} - \boldsymbol{\beta}^0_{sr})\right].
\end{align}

\subsubsection{Weights}
The conditional posterior distribution for $w_{sj}$ is given by:
\begin{align}
    p(w_{sj}\mid\ldots) &\propto \exp\left\{-w_{sj}\right\}
    \frac{[\beta_s^w]^{\alpha_s^w}}{\Gamma(\alpha_s^w)} [w_{sj}]^{\alpha_s^w - 1}\exp\left\{-\beta_s^w w_{sj}\right\}
    \left[\prod_{i=1}^K \prod_{r=1}^R [w_{sj}]^{z_{sijr}}\right], \\
    &\propto [w_{sj}]^{\sum_i\sum_r z_{sijr} + \alpha_s^w - 1}\exp\left\{-(\beta_s^w + 1)\right\}, \\
    &= [w_{sj}]^{\sum_i m_{sij} + \alpha_s^w - 1}\exp\left\{-(\beta_s^w + 1)\right\},
\end{align}
which implies
\begin{equation}
    w_{sj} \mid \ldots \sim \text{Gamma}\left(\alpha_s^w + \sum_{i} m_{sij}, \beta_s^w + 1\right).
\end{equation}
The optimal factor for mutational burden parameter $w_{sj}$ is given by:
\begin{align}
 q^*\left(w_{sj}\right) &\ \Gamma\left(\theta_{sj}^{w\alpha},\theta_{sj}^{w\beta} \right), 
\end{align}
where
\begin{align}
    \theta_{sj}^{w\alpha} &= \mathbb{E}\left[\alpha_s^w\right] + \sum_{i} m_{sij}, \\
    \theta_{sj}^{w\beta} &= \mathbb{E}\left[\beta_s^w\right] + 1.
\end{align}

The conditional posterior distribution for $\alpha_s^w$ is given by:
\begin{align}
    \alpha_s^w \mid \ldots &\propto \lambda_s^w\exp\left\{-\lambda_s^w\alpha_s^w\right\}\left[\prod_{j=1}^{N_s} \frac{([\beta_{s}^w]^{\alpha_s^w})}{\Gamma(\alpha_s^w)}[w_{sj}]^{\alpha_s^w-1} \exp\left\{-\beta_s^w w_{sj}\right\}\right],
\end{align}
which is an unknown density. We can sample from the posterior of $\alpha_s^w$ using a manhattan step. The optimal factor for $\alpha_s^w$ is given by:
\begin{align}
    q^*\left(\alpha_s^w\right) &\propto  \lambda_s^w\exp\left\{-\lambda_s^w\alpha_s^w\right\}\frac{(\exp\left\{\mathbb{E}\left[\log \beta_{s}^w \right]\right\}^{N_s \alpha_s^w})}{\Gamma(\alpha_s^w)^{N_s}}\left[\prod_{j=1}^{N_s} \exp\left\{\mathbb{E}\left[\log w_{sj}\right]\right\}^{\alpha_s^w-1} \exp\left\{-\mathbb{E}\left[\beta_s^w\right] w_{sj}\right\}\right], 
\end{align}
which is also an unknown density. However, the expectation can be evaluated with the law of the unconscious statistician: 
\begin{align}
     \mathbb{E}_q\left[f(\alpha_s^w)\right] &= \int_{0}^{\infty}f(\alpha_s^w)\frac{1}{C}\lambda_s^w\exp\left\{-\lambda_s^w\alpha_s^w\right\}\frac{(\exp\left\{\mathbb{E}\left[\log \beta_{s}^w \right]\right\}^{N_s \alpha_s^w})}{\Gamma(\alpha_s^w)^{N_s}} \\
     &\qquad \times \left[\prod_{j=1}^{N_s} \exp\left\{\mathbb{E}\left[\log w_{sj}\right]\right\}^{\alpha_s^w-1} \exp\left\{-\mathbb{E}\left[\beta_s^w\right] w_{sj}\right\}\right]d\alpha_s^w, \label{w_shape_expect} \\
     C &= \int_{0}^{\infty}\lambda_s^w\exp\left\{-\lambda_s^w\alpha_s^w\right\}\frac{(\exp\left\{\mathbb{E}\left[\log \beta_{s}^w \right]\right\}^{N_s \alpha_s^w})}{\Gamma(\alpha_s^w)^{N_s}} \\
     &\qquad \times \left[\prod_{j=1}^{N_s} \exp\left\{\mathbb{E}\left[\log w_{sj}\right]\right\}^{\alpha_s^w-1} \exp\left\{-\mathbb{E}\left[\beta_s^w\right] w_{sj}\right\}\right]d\alpha_s^w. \label{w_shape_const}
\end{align}
In our implementation, numeric integration is performed via the \texttt{hcubature} function of the R package \texttt{cubature}~\citep{cubature_package}.

The conditional posterior distribution for $\beta_s^w$ is given by:
\begin{align}
    p(\beta_s^w\mid\ldots) &\propto \left[\prod_{j=1}^{N_s} \frac{([\beta_{s}^w]^{\alpha_s^w})}{\Gamma(\alpha_s^w)}[w_{sj}]^{\alpha_s^w-1} \exp\left\{-\beta_s^w w_{sj}\right\}\right][\beta_s^w]^{a_s^w-1}\exp\left\{-b_s^w\beta_s^w\right\}, \\
    &\propto [\beta_s^w]^{N_s\alpha_s^w + a_s^w -1}\exp\left\{-(\sum_{j=1}^{N_s}w_{sj}+b_s^w)\beta_s^w\right\},
\end{align}
which implies
\begin{equation}
    \beta_s^w\mid\ldots \sim \text{Gamma}\left(N_s\alpha_s^w + a_s^w, \sum_{j=1}^{N_s} w_{sj}+b_s^w\right).
\end{equation}
The optimal factor for $\beta_s^w$ is given by:
\begin{align}
 \log q^*\left(\beta_s^w\right) &= \Gamma\left( \theta_{s}^{\beta 1}, \theta_{s}^{\beta 2} \right), 
\end{align}
where
\begin{align}
    \theta_{s}^{\beta 1} &= N_s\mathbb{E}\left[\alpha_s^w\right] + a_s^w,\\
    \theta_{s}^{\beta 2} &= \sum_{j=1}^{N_s} \mathbb{E}\left[w_{sj}\right]+b_s^w.
\end{align}

\newpage

\section{CAVI Algorithm}

Based on our work in Section~\ref{sec:supporting_calcs}, we can rewrite our mean field, including the parameters for each factor:
\begin{align}
    q(\boldsymbol{\theta}) &= \left[\prod_{r=1}^R q\left(\mathbf{P}_r\mid\boldsymbol{\theta}_{r}^p\right)\right]\left[\prod_{s=1}^S\prod_{j=1}^{N_s}q\left(\mathbf{E}_{sj}\mid \boldsymbol{\theta}_{sj}^e\right)\right] \left[\prod_{s=1}^S\prod_{j=1}^{N_s}q\left(w_{sj}\mid \theta_{sj}^{w\alpha}, \theta_{sj}^{w\beta}\right)\right] \\
    & \times \left[ \prod_{s=1}^S\prod_{j=1}^{N_s}\prod_{r=1}^R q\left(a_{sjr}\mid\theta_{srj}^a\right)q\left(a^*_{sjr}\mid \theta_{srj}^{a*}\right)\right] \left[ \prod_{s=1}^S\prod_{r=1}^R q\left(\boldsymbol{\beta}_{sr}\mid\boldsymbol{\theta}_{sr}^{\beta\mu},\boldsymbol{\theta}_{sr}^{\beta\Sigma}\right)q\left(\tau_{sr}\mid \theta_{sr}^{\tau\alpha
    }, \theta_{sr}^{\tau\beta
    }\right) \right]\\
    & \times \left[ \prod_{s=1}^S q(\alpha_s^w) q(\beta_s^w\mid \theta_{s}^{\beta 1}, \theta_{s}^{\beta 2})  \right].
\end{align}
The full CAVI algorithm is as follows:
\begin{breakablealgorithm}
    \setstretch{1.2}
    \small
    \begin{flushleft}
    \caption{CAVI for BAP Multi-NMF}\label{alg:CAVI}
    $\bf{Input}$: Data $\left\{\mathbf{M}_s\right\}$
    
    $\bf{Initialize}$: Set initial variational parameters $\boldsymbol{\theta}(0)$. 
    
    \textbf{For $t=1,\ldots$ }
    \begin{enumerate}
        \item For $s=1,\dots,S$, $i = 1,\dots, K$, $j=1,\dots,N_s$, and $r=1,\dots,R$, set:
        \begin{align*}
            z_{sijr} &= m_{sij} \frac{p_{ir}e_{sjr}}{\sum_{r} p_{ir}e_{sjr}}.
        \end{align*}
        \item For $s=1,\dots, S$, $j=1,\dots,N_s$, set:
        \begin{align*}
            \theta_{sj}^{w\alpha} &= \mathcal{E}\left[\alpha_{s}^w\right] + \sum_{i=1}^{N_s} m_{sij}, \\
            \theta_{sj}^{w\beta} &= \frac{\theta_s^{\beta 1}}{\theta_s^{\beta 2}} + 1.
        \end{align*}
        \item For $s=1,\dots, S$, set:
        \begin{align*}
            q(\alpha_s^w) &= \frac{1}{C}\lambda_s^w\exp\left\{-\lambda_s^w\alpha_s^w\right\}\frac{(\exp\left\{\psi(\theta_s^{\beta 1}) - \log \theta_s^{\beta 2}\right\}^{N_s \alpha_s^w})}{\Gamma(\alpha_s^w)^{N_s}}\left[\prod_{j=1}^{N_s} \exp\left\{\psi\left( \theta_{sj}^{w\alpha}\right)-\log  \theta_{sj}^{w\beta}\right\}^{\alpha_s^w-1} \exp\left\{-\mathbb{E}\left[\beta_s^w\right] w_{sj}\right\}\right], 
        \end{align*}
        where:
        \begin{align*}
            C &= \int_{0}^{\infty}\lambda_s^w\exp\left\{-\lambda_s^w\alpha_s^w\right\}\frac{(\exp\left\{\psi(\theta_s^{\beta 1}) - \log \theta_s^{\beta 2}\right\}^{N_s \alpha_s^w})}{\Gamma(\alpha_s^w)^{N_s}}\left[\prod_{j=1}^{N_s} \exp\left\{\psi\left( \theta_{sj}^{w\alpha}\right)-\log  \theta_{sj}^{w\beta}\right\}^{\alpha_s^w-1} \exp\left\{-\mathbb{E}\left[\beta_s^w\right] w_{sj}\right\}\right]d\alpha_s^w,
        \end{align*}
        and $\psi(\cdot)$ is the digamma function.
        \item For $s=1,\dots, S$, set:
        \begin{align*}
            \theta_{s}^{\beta 1} &= N_s\mathbb{E}\left[\alpha_s^w\right] + a_s^w,\\
            \theta_{s}^{\beta 2} &= \sum_{j=1}^{N_s} \frac{\theta_{sj}^{w\alpha}}{\theta_{sj}^{w\beta}}+b_s^w.
        \end{align*}
        \item For $i=1,\dots,K$, $r=1,\dots, R$, set:
        \begin{align*}
            \theta_{ir}^p &= \alpha_{ir}^p + \sum_s \sum_j z_{sijr}.
        \end{align*}
        \item For $s=1,\dots,S$, $j=1,\dots,N_s$, set:
        \begin{align*}
        \theta_{srj}^e &= \theta^a_{srj}\alpha_s^{e1} + (1- \theta^a_{srj})\alpha_s^{e0} + \sum_{k=1}^K z_{sijr}.
        \end{align*}
        \item For $s=1,\dots,S$, $j=1,\dots,N_s$, $r=1,\dots,R$, set:
        \begin{align*}
            \text{logit}\left( \theta^a_{srj} \right) &= \log \Gamma \left(\alpha_s^{e1} + \sum_{t \neq r} \theta_{srj}^a \alpha_s^{e1} + (1-\theta_{srj}^a) \alpha_s^{e0}\right) \\
            & + \frac{1}{2} \psi^{1}\left(\alpha_s^{e1} + \sum_{t \neq r} \theta_{srj}^a \alpha_s^{e1} + (1-\theta_{srj}^a) \alpha_s^{e0}\right)\left( \sum_{t \neq r} \left( \alpha_s^{e1} -  \alpha_s^{e0}\right)^2 \theta_{srj}^a \left(1-\theta_{srj}^a\right)\right) \\
            & - \log \Gamma\left(\alpha_s^{e1}\right) +  \left(\alpha_s^{e1}-1\right) \left( \psi(\theta_{srj}) - \psi\left(\sum_r \theta_{srj}\right) \right) \\
            & + \log \left( 1 - \left(1 + exp\left\{-c\right\}\right)^{-1}\right) - 1/2 * \boldsymbol{x}_{sj}^\intercal \boldsymbol{\theta}_{sr}^{\beta\Sigma}\boldsymbol{x}_{sj} \\
            &\times \frac{e^{c}\left(-3a \left(\left[\boldsymbol{\theta}^{\beta\mu}_{sr}\right]^\intercal\boldsymbol{x}_{sj}\right)\left(2e^{c} -3a\left(\left[\boldsymbol{\theta}^{\beta\mu}_{sr}\right]^\intercal\boldsymbol{x}_{sj}\right)^3 + 2 \right) - 6a\left(\left[\boldsymbol{\theta}^{\beta\mu}_{sr}\right]^\intercal\boldsymbol{x}_{sj}\right) + b^2\right)}{\left(e^c+1\right)^2}  \\
            &  -  \log \Gamma \left(\alpha_s^{e0} + \sum_{t \neq r} \theta_{srj}^a \alpha_s^{e1} + (1-\theta_{srj}^a) \alpha_s^{e0}\right) \\
            & + \frac{1}{2} \psi^{1}\left(\alpha_s^{e0} + \sum_{t \neq r} \theta_{srj}^a \alpha_s^{e1} + (1-\theta_{srj}^a) \alpha_s^{e0}\right)\left( \sum_{t \neq r} \left( \alpha_s^{e1} -  \alpha_s^{e0}\right)^2 \theta_{srj}^a \left(1-\theta_{srj}^a\right)\right)\\
            & + \log \Gamma\left(\alpha_s^{e0}\right) -  \left(\alpha_s^{e0}-1\right)  \left( \psi(\theta_{srj}) - \psi\left(\sum_r \theta_{srj}\right) \right)\\
            & -  \log \left( \left(1 + exp\left\{-c\right\}\right)^{-1}\right) + 1/2 * \boldsymbol{x}_{sj}^\intercal \boldsymbol{\theta}_{sr}^{\beta\Sigma}\boldsymbol{x}_{sj} \\
            &\times \frac{b^2e^c - 3a\left(-\left[\boldsymbol{\theta}^{\beta\mu}_{sr}\right]^\intercal\boldsymbol{x}_{sj}\right)\left(\left(3a\left(-\left[\boldsymbol{\theta}^{\beta\mu}_{sr}\right]^\intercal\boldsymbol{x}_{sj}\right)^3 - 2\right)e^c - 2\right) + 6ab\left(-\left[\boldsymbol{\theta}^{\beta\mu}_{sr}\right]^\intercal\boldsymbol{x}_{sj}\right)^2e^c}{\left(e^c + 1\right)^2},
        \end{align*}
        where: 
        \begin{align*}
            a &= 0.07056, \\
            b &= 1.5976, \\
            c & = \left(a * \left(-\left[\boldsymbol{\theta}^{\beta\mu}_{sr}\right]^\intercal\boldsymbol{x}_{sj}\right)^3 + b * \left(-\left[\boldsymbol{\theta}^{\beta\mu}_{sr}\right]^\intercal\boldsymbol{x}_{sj}\right)\right),
        \end{align*}
        and $\psi^1(\cdot)$ is the first derivative of the digamma function.
        \item For $s=1,\dots,S$, $j=1,\dots,N_s$, $r=1,\dots,R$, set:
        \begin{align*}
            \theta_{srj}^{a*} &= \left[\boldsymbol{\mu}_{sr}^{\beta}\right]^\intercal\boldsymbol{x}_{sj}.
        \end{align*}
        \item For $s=1,\dots,S$, $r=1,\dots,R$, set:
        \begin{align*}
            \boldsymbol{\theta}_{sr}^{\beta\Sigma} &= \left(\frac{\theta_{sr}^{\tau\alpha }}{\theta_{sr}^{\tau\beta }}\mathcal{I}_{Q_s} + \mathbf{X}_s^\intercal \mathbf{X}_s\right)^{-1}, \\
            \boldsymbol{\theta}_{sr}^{\beta\mu} &= \boldsymbol{\Sigma}_{sr}^{\beta}\left(\frac{\theta_{sr}^{\tau\alpha }}{\theta_{sr}^{\tau\beta }}\mathcal{I}_{Q_s}\boldsymbol{\beta}^0_{sr} + \mathbf{X}_s^\intercal\mathbf{\Theta}^{a*}_{sr}\right),
        \end{align*}
        where $\mathbf{\Theta}^{a*}_{sr} = (\theta_{sr1}^{a*}, \dots, \theta_{srN_s}^{a*})^\intercal$.
        \item For $s=1,\dots,S$, $r=1,\dots,R$, set:
        \begin{align*}
             \theta_{sr}^{\tau\alpha } &=  \gamma_1 + Q_s/2, \\
             \theta_{sr}^{\tau\beta} &= \gamma_2 + \frac{1}{2}\left(\text{Tr}\left(\boldsymbol{\theta}_{sr}^{\beta\Sigma}\right) + \left(\boldsymbol{\theta}_{sr}^{\beta\mu} - \boldsymbol{\beta}_{sr}^0\right)^{\intercal} \left(\boldsymbol{\theta}_{sr}^{\beta\mu} - \boldsymbol{\beta}_{sr}^0\right)\right) .
        \end{align*}
    \end{enumerate}
    \textbf{End for}
        
    \bf{Output:} variational approximation $q^*(\boldsymbol{\theta})$ 
    \end{flushleft}
\end{breakablealgorithm}

\subsection{Initialization Strategy}

In this section, we provide the overview of the initialization strategies implemented in the R package implementation of CAVI for the BAP Multi-NMF. The algorithm's performance is highly sensitive to the choice of initialization, as the quality of variational approximations  can significantly depend on the initial values. In Section~\ref{sec:discovery-initalization}, we discuss the initialization strategy designed for estimating de novo mutational signatures. In Section~\ref{sec:recovery-discovery-initalization}, we explain how to initialize BAP Multi-NMF when the goal is to recover previously identified signatures in the analysis.

\subsubsection{Discovery Only Version}\label{sec:discovery-initalization}

In the discovery-only version of BAP Multi-NMF, the \texttt{R} argument specifies the maximum rank of mutational signatures matrix, $\mathbf{P}$.

The procedure is as follows:
\begin{enumerate}
    \item  Estimate the initial signatures and exposures matrices, $\mathbf{P}(0)$, $\left\{\mathbf{E}_s(0)\right\}$ using a frequentist NMF approach. We use the standard NMF algorithm provided by the \texttt{NMF} R package with default parameters, setting the rank to the user input \texttt{R}~\citep{NMF_package}.
    \item For $i=2,\dots,R$: discard $\mathbf{P}(0)_i$ if $\text{max } \text{Cos}\left(\mathbf{P}(0)_i, \mathbf{P}(0)_{1:(i-1)}\right) > 0.5$.
    \item Set $R'$ equal to the number of columns remaining in $\mathbf{P}(0)$. Recalculate $\mathbf{P}(0)'$, $\left\{\mathbf{E}_s(0)'\right\}$ using the frequentist NMF adjusting the rank to  $R'$.
    \item Randomly initialize $R - R'$ mutational signatures using dirichlet draws with the prior concentration. This step generates $\mathbf{P}(0)"$, $\left\{\mathbf{E}_s(0)"\right\}$.
    \item Combine the recalibrated and randomly initialized sets of signatures and exposures to obtain the final initial values:
    \begin{align*}
        \mathbf{P}(0) &= \left(\mathbf{P}(0)', \mathbf{P}(0)"\right) \\
        \mathbf{E}_s(0) &= \left(\mathbf{E}_s(0)', \mathbf{E}_s(0)"\right)
    \end{align*}
    \item Begin CAVI, starting at step 1 with $\mathbb{E}[p_{ir}]=\mathbf{P}(0)_{ir}$, $\mathbb{E}[e_{srj}]=\mathbf{E}(0)_{srj}$, and all other expectations set according to the prior distribution.
\end{enumerate}

\subsubsection{Recovery-Discovery Version}\label{sec:recovery-discovery-initalization}
In the recovery-discovery version of BAP Multi-NMF, the \texttt{R} argument is interpreted as the maximum number of discovered signatures. The initial exposures for the recovered signatures are calculated via frequentist NMF conditional on the recovered signatures matrix $\mathbf{P}_r$, and the both the initial discovered signatures and exposures are randomly initialized via draws from the prior distribution.

\newpage

\section{Additional Simulation Results}

\begin{figure}[!htb]
    \centering
    \includegraphics[width=0.5\linewidth]{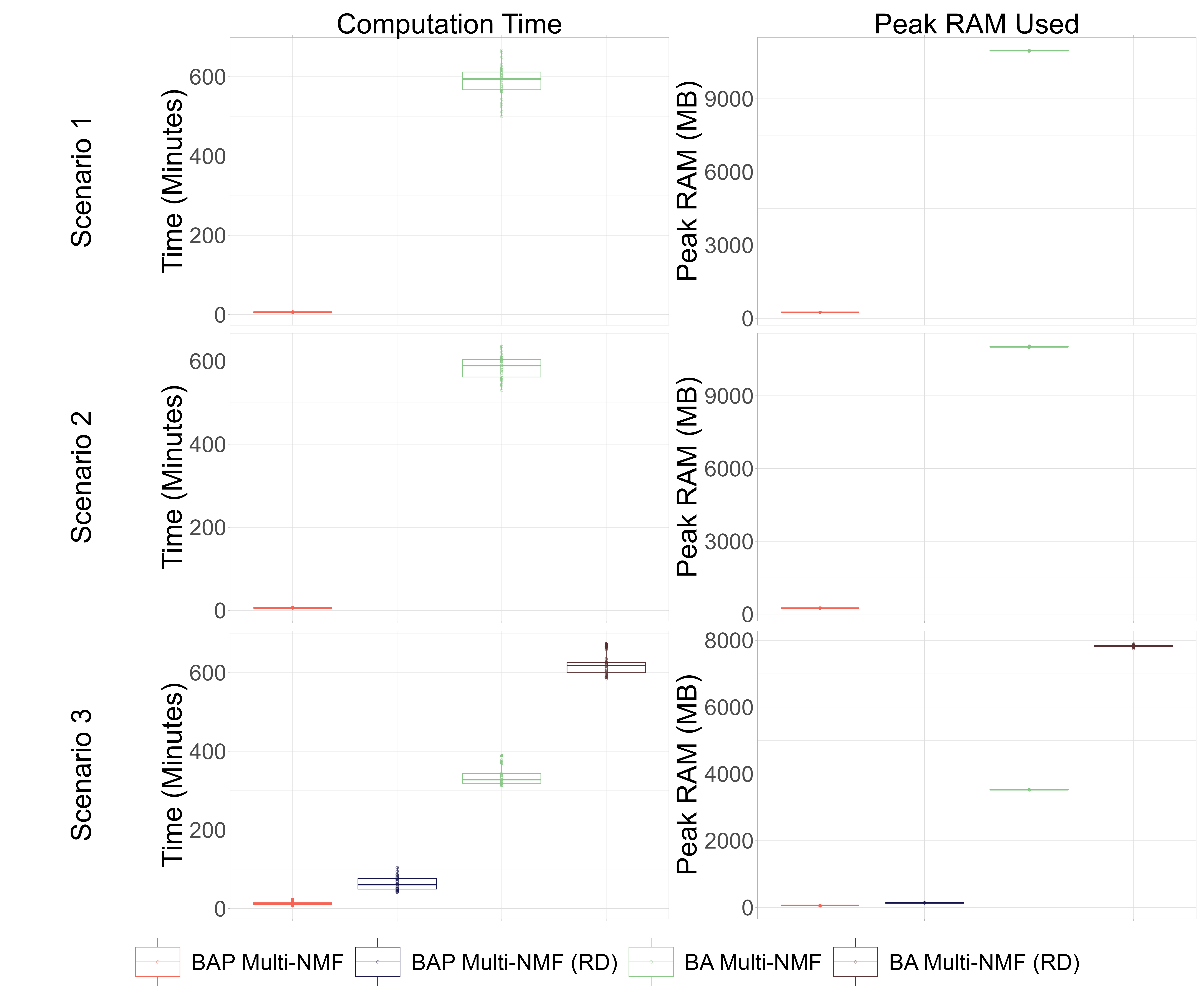}
    \caption{Computational resources required to run BAP Multi-NMF and BA Multi-NMF across the three simulation scenarios. The first column visualizes the computational time, in minutes, required to run each method, while the second column shows the computational memory used, in MB of RAM. These results highlight the significant computational efficiency of BAP Multi-NMF, particularly in scenarios with larger datasets.}
    \label{fig:sim_settings}
\end{figure}

\newpage

\section{Additional PCAWG Analysis Results}

\begin{figure}[!htb]
    \centering
    \includegraphics[width=\linewidth]{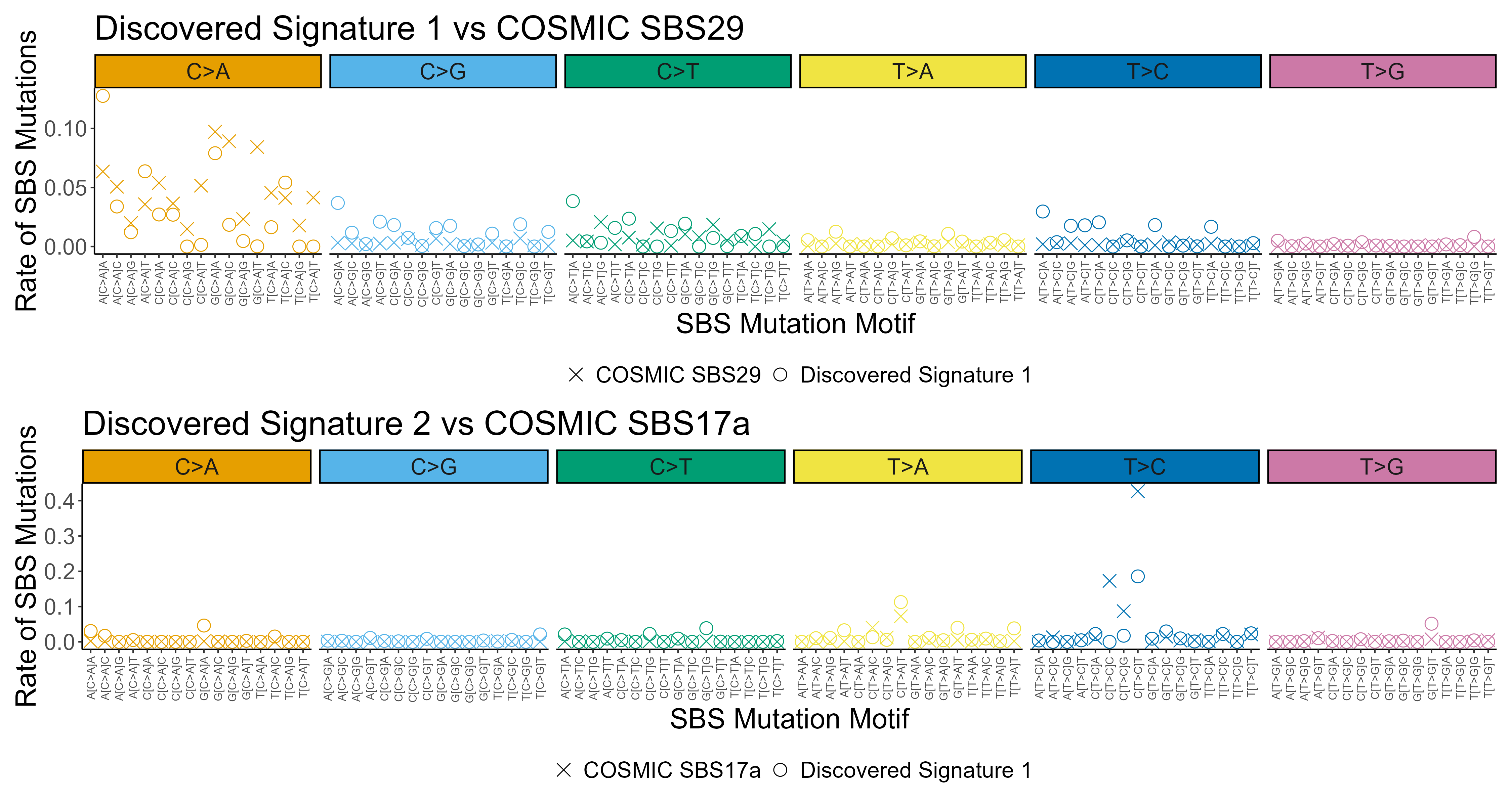}
    \caption{Comparisons of Discovered Signatures 1 and 2 to their closest COSMIC matches, SBS29 and SBS17a, respectively. Values shown are representative of the frequencies of the 96 SBS mutation categories associated with each mutational signature.}
    \label{fig:d_sigs}
\end{figure}

\begin{figure}[!htb]
    \centering
    \includegraphics[width=0.75\linewidth]{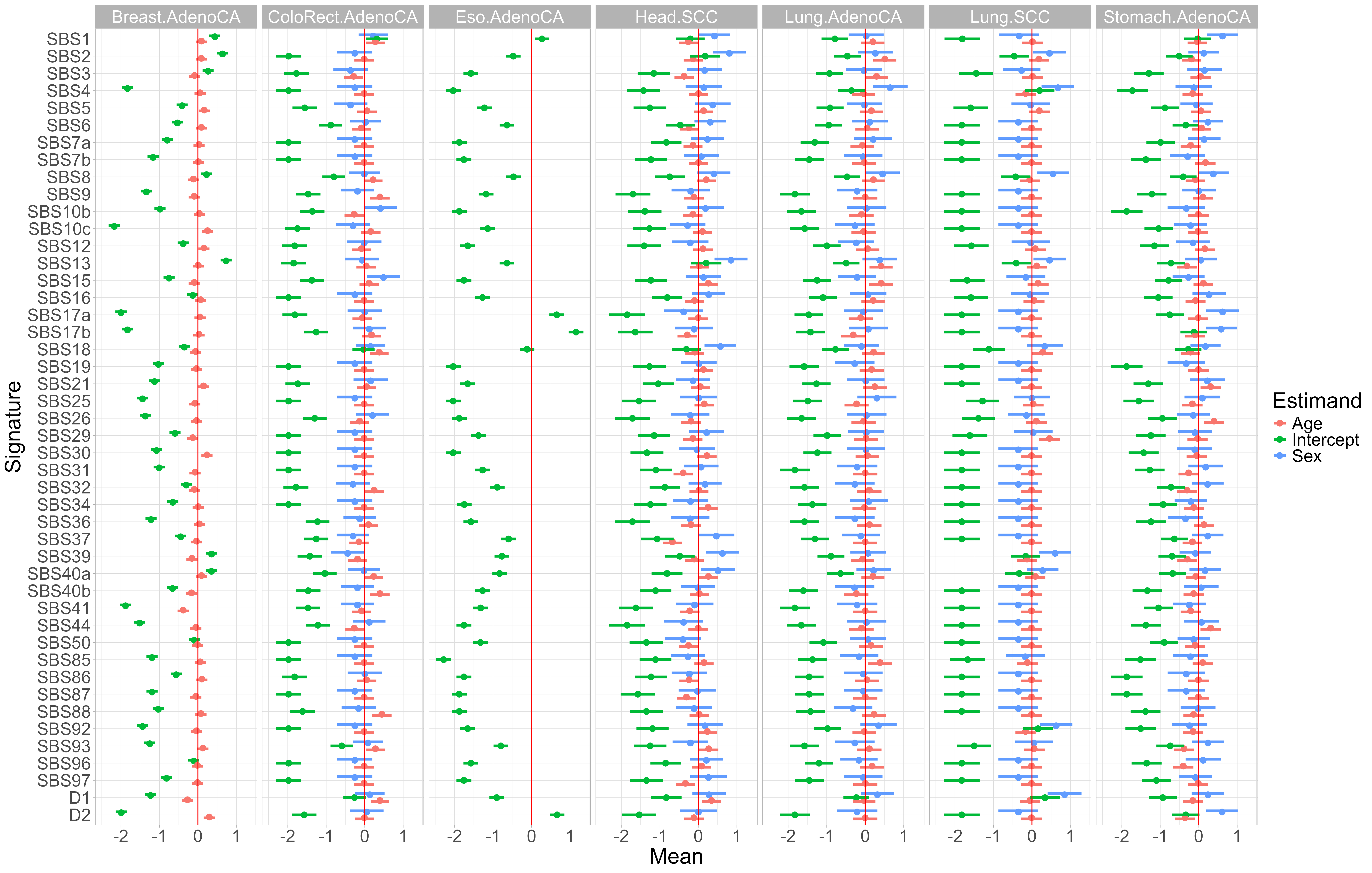}
    \caption{Visualization of study-specific covariate effects, reported as a 95\% credible interval based on the variational approximation of BAP Multi-NMF. Due to limitations noted in Section~2, we excluded the sex covariate for the breast adenocarcinoma samples, and we excluded both covariates for esophagus adenocarcinoma.}
    \label{fig:pcawg_covariates}
\end{figure}

\newpage

\bibliographystyle{apalike} 
\bibliography{refs}       